\documentclass{ws-ijgmmp}
\usepackage[utf8]{inputenc}

\usepackage{xcolor}
\usepackage{amsmath, array, amssymb, amsfonts, mathtools}
\usepackage{xfrac}
\usepackage[inline]{enumitem}

\usepackage[english]{babel}
\usepackage[all]{xy}

\usepackage{lmodern}

\usepackage{booktabs}
\usepackage{caption}
\usepackage{dsfont}
\usepackage{mathtools}

\usepackage[hidelinks]{hyperref}

\usepackage{textcomp}

\usepackage{multicol}
\usepackage[square,numbers,sort,compress,semicolon,merge]{natbib}
\let\cite\citep 
\usepackage{hyperref}
\hypersetup{colorlinks=true, urlcolor=blue, citecolor=blue, linktoc=page}

\makeatletter
\renewcommand*\env@matrix[1][\arraystretch]{%
  \edef\arraystretch{#1}%
  \hskip -\arraycolsep
  \let\@ifnextchar\new@ifnextchar
  \array{*\c@MaxMatrixCols c}}
\makeatother

\renewcommand\P{\mathcal{P}}
\newcommand\M{\mathcal{M}}

\newcommand\RR{\mathbb{R}}
\newcommand\CC{\mathbb{C}}
\renewcommand\1{\textbf{1}}
\newcommand\id{\textit{id}}
\newcommand\T{\mathcal{T}}

\renewcommand\H{\mathcal{H}}

\renewcommand\S{\mathcal{S}}

\newcommand\SU{\mathcal{SU}}
\newcommand\U{\mathcal{U}}
\newcommand\SO{\mathcal{SO}}
\newcommand\SL{\mathcal{SL}}

\newcommand\K{\mathcal{K}}
\newcommand\J{\mathcal{J}}

\newcommand\GL{\mathcal{GL}}

\newcommand\W{\mathcal{W}}
\newcommand\vphi{\varphi}

\renewcommand\l{\text{\tiny{L}}}
\newcommand\ww{\text{\tiny{W}}}
\renewcommand\ss{\text{s}}
\newcommand\n{\text{\tiny{N}}}
\newcommand\sP{\mathsf{P}}
\newcommand\sC{\mathsf{C}}
\newcommand\sT{\mathsf{T}}
\newcommand\sW{\mathsf{W}}
\newcommand\sS{\mathsf{S}}
\newcommand\sZ{\mathsf{Z}}

\newcommand\sE{\mathsf{E}}
\newcommand\sR{\mathsf{R}}
\newcommand\sG{\mathsf{G}}

\renewcommand\epsilon{\varepsilon}

\newcommand\rarrow{\rightarrow}

\newcommand*\vect[1]{\begin{pmatrix}#1 \end{pmatrix}}

\newcommand\LieG{\mathfrak{g}}
\newcommand\LieH{\mathfrak{h}}

\newcommand\su{\mathfrak{su}}
\newcommand\so{\mathfrak{so}}
\renewcommand\sl{\mathfrak{sl}}
\newcommand\co{\mathfrak{co}}
\newcommand\gl{\mathfrak{gl}}
\newcommand\LieK{\mathfrak{k}}
\newcommand\LieJ{\mathfrak{j}}

\renewcommand\t{\widetilde}
\newcommand\h{\widehat}
\renewcommand\b{\bar }
\newcommand\w{\wedge}
\renewcommand\d{\partial}
\newcommand\s{\sigma}

\renewcommand\-{^{-1}}
\newcommand\Ad{\text{Ad}}

\renewcommand\id{\text{id}}
\renewcommand\1{\mathds{1}}

\renewcommand{\descriptionlabel}[1]{\textit{#1}}

\DeclareMathOperator{\Aut}{Aut}
\DeclareMathOperator{\Tr}{Tr}
\DeclareMathOperator{\vol}{vol}
\DeclareMathOperator{\Spin}{Spin}
\DeclareMathOperator{\Herm}{Herm}

\DeclarePairedDelimiter\norm{\lVert}{\rVert}

\allowdisplaybreaks

\newtheorem{thm}{Theorem}

\newtheorem{cor}[thm]{Corollary}
\newtheorem{prop}[thm]{Proposition}

\hypersetup{
pdfauthor={J. Attard, J. François, S. Lazzarini, T. Masson},
pdftitle={The dressing field method of gauge symmetry reduction, a review with examples},
pdfsubject={},
pdfcreator={pdflatex},
pdfproducer={pdflatex},
pdfkeywords={}
}

\usepackage{xcolor}
\newcounter{mnotecount}[section]
\renewcommand{\themnotecount}{\thesection.\arabic{mnotecount}}
\newcommand{\mnote}[1]
{\protect{\stepcounter{mnotecount}}$^{\mbox{\footnotesize
$
\bullet$\themnotecount}}$ \marginpar{
\raggedright\tiny\em
$\!\!\!\!\!\!\,\bullet$\themnotecount: #1} }

\begin{document}

\markboth{J. Attard, J. François, S. Lazzarini,T. Masson}
{The dressing field method of gauge symmetry reduction, a review with examples}

%
\catchline{}{}{}{}{}
%

\title{The dressing field method of gauge symmetry reduction, \\ a review with examples}

\author{J. Attard, J. François, S. Lazzarini, T. Masson}

\address{Aix Marseille Univ, Université de Toulon, CNRS, CPT, Marseille, France}

\maketitle


\begin{abstract}
Gauge symmetries are a cornerstone of modern physics but they come with technical difficulties when it comes to quantization, to accurately describe particles phenomenology or to extract observables in general. These shortcomings must be met by essentially finding a way to effectively reduce gauge symmetries. We propose a review of a way to do so which we call the dressing field method. We show how the BRST algebra satisfied by gauge fields, encoding their gauge transformations, is modified. We outline noticeable applications of the method, such as the electroweak sector of the Standard Model and the local twistors of Penrose. 
\end{abstract}

\keywords{Gauge field theories; Dressing Field Method; BRST algebra; symmetry reduction; electroweak theory; conformal geometry; Cartan geometry; tractors; twistors.}

\vspace{1mm}

PACS numbers:  02.40.Hw, 11.15.-q,11.25.Hf.

\vspace{1.5mm}


\tableofcontents

\markboth{J. Attard, J. François, S. Lazzarini,T. Masson}
{The dressing field method of gauge symmetry reduction, a review with examples}

\section{Introduction}  

To this day, modern Field Theory framework (either classical or quantum), so successful in describing Nature from elementary particles to cosmology, rests on few keystones, one of which being the notion of gauge symmetry. Elementary fields are subject to local transformations which are required to leave invariant the theory (the Lagrangian). These transformations thus form a so-called local symmetry of the theory, known as gauge symmetry. This notion, originates with Weyl's $1918$ unified theory resting on the  idea of local \emph{scale}, or \emph{gauge}, invariance \cite{Weyl1918, Weyl1919, ORaif1997}. The heuristic appeal of gauge symmetries is that imposing them on a theory of free fields requires, a minima, the introduction of fundamental interactions through minimal coupling.
This is the content of the so-called \emph{gauge principle} for Field Theory,\footnote{See \cite{Martin2002} for a critical discussion of its scope and limits.} captured by Yang's well-known aphorism: ``symmetry dictates interaction'' \cite{YangSelPap}.\footnote{ Weyl  topped this with an even stronger endorsement of the importance of symmetries in physics: ``As far as I see, all a priori statements in physics have their origin in symmetry'' \cite{Weyl1952}.} 

This is one of the major conceptual breakthrough of the century separating us from Hilbert's lectures on the foundations of mathematics and physics. And the story of the interactions between gauge theories and differential geometry is a highlight of the long history of synergy between mathematics and physics.\footnote{See the nice short appendix by S. S. Chern of the book on differential geometry he co-authored \cite{Chern-Chen-Lam1999}.}

\smallskip
In spite of their great theoretical appeal, gauge theories come with some shortcomings. Prima facie indeed,  gauge symmetries forbid mass terms for (at least) the interaction fields, which was known to be in contradiction with the phenomenology of the nuclear interactions. Also, the quantization of gauge theories via Feynman's path integral has its specific problem because integrating on gauge equivalent fields configurations makes it ill-defined. Finally, it is in general not so straightforward to extract observables from a gauge theory since the physical content must be gauge invariant, e.g the abelian (Maxwell-Faraday) field strength or Wilson loops. An issue made acutely clear in General Relativity (GR), where observables must be diffeomorphic invariant. Addressing these shortcomings essentially boils down to finding a way to \emph{reduce effectively gauge symmetries}, in part or completely. One can think of only a few ways to do so, among which we mention the three most prominent. 

First, gauge fixing: one selects a representative in the gauge orbit of each gauge field. This is usually the main approach followed to make contact with physical predictions: one only needs to make sure that the results are independent of the choice of gauge. This is also how a sensible quantization procedure is carried on, for example through the Fadeev-Popov procedure. However, a consistent choice of gauge might not necessarily be possible in all circumstances, a fact known as the \emph{Gribov ambiguity} \cite{Gribov, Singer}.

Second, one can try to implement a spontaneous symmetry breaking mechanism (SSBM). This is famously known to be the standard interpretation of the Brout-Englert-Higgs-Guralnik-Hagen-Kibble (BEHGHK) mechanism \cite{Englert-Brout, Higgs64, Guralnik-Hagen-Kibble}, which historically solved the masses problem for the weak gauge bosons in the electroweak unification of Glashow-Weinberg-Salam, and by extension of the masses of particles in the Standard Model of Particles Physics. We stress that this interpretation presupposes settled the philosophical problem of the ontological status of gauge symmetries: by affirming that a gauge symmetry can be ``spontaneously broken'', one states that it is a structural feature of reality rather than of our description of it. While this remains quite controversial in philosophy of physics, given  the empirical success of the BEHGHK mechanism, a pragmatic mind could consider the debate closed via an inference to the best explanation. We will show here that this conclusion would be hasty.  

 Finally, one can seek to apply the bundle reduction theorem. This is a result of the fiber bundle theory, widely known to be the geometric underpinning of gauge theories, stating the circumstances under which a bundle with a given structure group can be reduced to a subbundle with a smaller structure group. Some authors have recast the BEHGHK mechanism in light of this theorem \cite{Trautman, Westenholz, Sternberg}.

\smallskip
In this paper we propose a brief review of another way to perform gauge symmetry reduction which we call the \emph{dressing field method}. It is formalized in the language of the differential geometry of fiber bundles and it has a corresponding BRST differential algebraic formulation. The method boils down to the identification of a suitable field in the geometrical setting of a gauge theory that allows to construct partially of fully gauge invariant variables out of the standard gauge fields. This formalizes and unifies several works and approaches encountered in past and recent literature on gauge theories, whose ancestry can be traced back to Dirac's pioneering ideas \cite{Dirac55, Dirac58}. 

The paper is thus organized.
 In Section~\ref{Reduction of gauge symmetries: the dressing field method} we outline the method and state the most interesting results (pointing to the published literature for proofs), one of which being the noticeable fact that the method allows to highlight the existence of gauge fields of non-standard kind; meaning that these implement the gauge principle but are not of the same geometric nature than the objects usually encountered in Yang-Mills theory for instance. 

In Sections~\ref{The electroweak sector of the Standard Model} and \ref{From tetrad to metric formulation of General Relativity} we illustrate the scheme by showing how it is applied to the electroweak sector of the standard model and GR. We  argue in particular that our treatment provides an alternative interpretation of the BEHGHK mechanism that is more in line with the conclusions of the community of philosophers of physics. 

In Section~\ref{Conformal Cartan geometry, tractors and twistors} we address the substantial example of the conformal Cartan bundle $\P(\M, H)$ and connection $\varpi$. Standard formulations of so-called Tractors and Twistors can then be found by applying the dressing field method to this geometry. Furthermore, from this viewpoint they appear to be clear instances of gauge fields of the non-standard kind alluded to above. This fact, as far as we know, has not been recognized.

In our conclusion, Section~\ref{Conclusion}, we indicate other possible applications of the method and stress the obvious  remaining open questions to be addressed.

\section{Reduction of gauge symmetries: the dressing field method}   
\label{Reduction of gauge symmetries: the dressing field method}   

As we have stated, the differential geometry of fiber bundles supplemented by the BRST differential algebra  are the mathematical underpinning of classical gauge theories. So, this is the language in which we will formalize our approach. Complementary material and detailed proofs can be found in \cite{GaugeInvCompFields, Francois2014, FLM2015_II, Attard-Francois2016_I}.

Let us give the main philosophy of the dressing field method in few words. From a mathematical point of view, a gauge field theory requires some spaces of fields on which the gauge group acts in a definite way. So, to define a gauge field theory, the spaces of fields themselves are not sufficient: one has to specify the actions of the gauge group on them. This implies that the same mathematical space can be considered as a space of different fields, according to the possible actions of the gauge group. 

Generally, the action is related to the way the space of fields is constructed. For instance, in the usual geometrical framework of gauge field theories, the primary structure is a principal fiber bundle $\P$, and the gauge group is its group of vertical automorphisms. Then, the sections of an associated vector bundle to $\P$, constructed using the action $\rho$ of the structure group on a vector space $V$, support an action of the gauge group which is directly related to $\rho$. 

The physical properties of a gauge field theory are generally encoded into a Lagrangian $L$ written in terms of the gauge fields (and their derivatives): it is required to be invariant when the gauge group acts on all the fields involved in its writing.

The main idea behind the dressing field method is to exhibit a very special field (the dressing field) \emph{out of the gauge fields in the theory}, with a specific gauge action. Then, one performs some change of field variables, very similar to some change of variables in ordinary geometry, by \emph{combining} in a convenient way (through sums and products \emph{when they make sense}) the gauge fields with the dressing field. The resulting “dressed fields” (new fields variables of the theory) are then subject to new actions of the gauge group, \emph{that can be deduced from the combination of fields}. In favorable situations, these dressed fields are invariant under the action of a subgroup of the gauge group: a part of the gauge group does not act anymore on the new fields of the theory, that is, the symmetry has been reduced.

Notice some important facts. Firstly, the dressed fields do not necessarily belong to the original space of fields from which they are constructed. Secondly, in general the dressing (\textit{i.e.} the combination of the dressing field with a field from the theory) looks like a gauge transformation. But we insist on the fact that it is not a gauge transformation. Finally, the choice of the dressing field relies sometimes on the physical content of the theory, that is on the specific form of the Lagrangian. So, the dressing field method can depend on the mathematical, as well as on the physical content of the theory.

Les us now describe the mathematical principles of the method.

\subsection{Composite fields}  
\label{Composite fields}  

Let $\P(\M, H)$ be a principal bundle over a manifold $\M$ equipped with a connection $\omega$ with curvature $\Omega$, and let $\vphi$ be a $\rho$-equivariant $V$-valued map on $\P$ to be considered as a section of the associated vector bundle $E=\P \times_H V$.

The group of vertical automorphisms of $\P$, 
\begin{equation*}
\Aut_v(\P):=\left\{\Phi:\P \rarrow \P \mid  \forall h\in H, \forall p \in \P,  \Phi(ph)=\Phi(p)h \text{ and } \pi \circ \Phi= \Phi \right\}
\end{equation*}
 is isomorphic to the gauge group $\H:=\left\{ \gamma :\P \rarrow H\mid  R^*_h\gamma(p)=h\- \gamma(p) h  \right\}$, the isomorphism being $\Phi(p)=p\gamma(p)$. The composition law of $\Aut_v(\P)$, $\Phi_1 \circ \Phi_2$, corresponds to the product $\gamma_1 \gamma_2$. 
 
 In this geometrical settings, the gauge group $\H \simeq \Aut_v(\P)$ acts on fields via pull-backs. It acts on itself as $\eta^{\gamma}:=\Phi^*\eta =\gamma^{-1} \eta \gamma$, and on connections $\omega$, curvatures $\Omega$ and $(\rho, V)$-tensorial forms $\vphi$ as,
\begin{align}
\label{ActiveGT}
\omega^\gamma &:=\Phi^*\omega=\gamma\-\omega\gamma + \gamma\- d\gamma,
&
\vphi^\gamma &:= \Phi^*\vphi=\rho(\gamma\-)\vphi,
\\ 
\Omega^\gamma &:=\Phi^*\Omega=\gamma\-\Omega \gamma, 
&
(D\vphi)^\gamma &:=\Phi^*D\vphi=D^\gamma \vphi^\gamma=\rho(\gamma\-)D\vphi.\notag
\end{align}
These are \emph{active} gauge transformations, formally identical but to be conceptually distinguished from \emph{passive} gauge transformations relating two local descriptions of the same global objects in local trivializations of the fiber bundles described as follows. Given two local sections $\s_1, \s_1$ of $\P$, related as  $\s_2=\s_1 h$, either over the same open set $\U$ of $\M$ or over the overlap of two open sets $\U_1 \cap \U_2$, one finds
\begin{align}
\label{PassiveGT}
\s_2^*\omega &=h\-\s_1^*\omega\  h + h\- dh, 
&
\s_2^*\vphi &= \rho(h\-)\s_1^*\vphi, 
\\
\s_2^*\Omega&=h\-\s_1^*\Omega\  h, 
&
\s_2^*D\vphi &=\rho(h\-)\s_1^*D\vphi.\notag
\end{align}
This distinction between active and passive gauge transformations is reminiscent of the distinction between diffeomorphism and coordinates transformations in GR. 
\smallskip

The main idea of the dressing field approach to gauge symmetries reduction is stated in the following

\begin{prop}[\cite{GaugeInvCompFields}] 
\label{P1}
Let $K$ and $G$ be subgroups of $H$ such that $K\subseteq G \subset H$. Note $\K\subset \H$  the gauge subgroup associated with $K$. Suppose there exists a map
\begin{align} 
\label{DF}
u:\P \rarrow G \quad \text{ satisfying the $K$-equivariance property }\quad  R_k^*u=k\-u.
\end{align}
Then this map $u$, that will be called a \emph{dressing field}, allows to construct through $f: \P \rarrow \P$ defined by $f(p)=pu(p)$, the following \emph{composite fields}
 \begin{align}
\label{CompFields}
\omega^u &:=f^*\omega=u\-\omega u+u\-du, 
&
\vphi^u &:=f^*\vphi= \rho(u\-)\vphi.
\end{align}
 which are $\K$-invariant and satisfy
\begin{align*}
\Omega^u &:=f^*\Omega=u\-\Omega u =d\omega^u+\tfrac{1}{2}[\omega^u, \omega^u],
\notag\\[1mm]
D^u\vphi^u &:=f^*D\vphi=\rho(u\-)D\vphi=d\vphi^u+\rho_*(\omega^u)\vphi^u.
\end{align*}
These composite field are $K$-horizontal and thus project on the quotient $\P/K$.
 \end{prop}

 The $\K$-invariance of the composite fields  \eqref{CompFields} is most readily proven. Indeed from the definition \eqref{DF} one has $f(pk)=f(p)$ so that  $f$ factorizes through a map $\P \rarrow \P/K$ and given $\Phi(p)=p\gamma(p)$ with $\gamma \in \K \subset \H$, one has $\Phi^*f^*=(f\circ \Phi)^*=f^*$. 

 The dressing field can be \emph{equally defined} by its $\K$-gauge transformation: $u^\gamma=\gamma\-u$, for any $\gamma \in \K\subset \H$. Indeed, given $\Phi$ associated to  $\gamma \in \K$ and  \eqref{DF} : $(u^\gamma)(p):=\Phi^*u(p)=u(\Phi(p))=u(p\gamma(p))=\gamma(p)^{-1}u(p)=(\gamma^{-1}u)(p)$.

Several comments are in order. First, \eqref{CompFields} looks \emph{algebraically} like \eqref{ActiveGT}: this makes easy to check algebraically that the composite fields are $\K$-invariant. Indeed, let $\chi \in \{\omega, \Omega, \vphi, \ldots\}$ denote a generic field  when performing an operation that applies equally well to any specific one. For two maps $\alpha, \alpha'$ with values in $H$, if one defines $\chi^\alpha$ \emph{algebraically} as in \eqref{ActiveGT}, then one has $(\chi^\alpha)^{\alpha'} = \chi^{\alpha \alpha'}$. This is for instance the usual way to compose the actions of two elements of the gauge group. But this relation is independent of the specific action of the gauge group on $\alpha$ and $\alpha'$, which could so belong to different spaces of representation of $\H$. Then $(\chi^u)^\gamma = (\chi^\gamma)^{u^\gamma} = (\chi^\gamma)^{\gamma\-u}=\chi^u$, where the last (and essential) equality is the one emphasized above.

Second, if $K=H$, then the composited fields \eqref{CompFields} are $\H$-invariant, the gauge symmetry is fully reduced, and they live on $\P/H \simeq \M$. This shows that the existence of a global dressing field is a strong constraint on the topology of the bundle $\P$: a $K$-dressing field means that the bundle is trivial along the $K$-subgroup, $\P \simeq \P/K \times K$, while a $H$-dressing field means its triviality, $\P\simeq \M \times H$ \cite[Prop.~2]{GaugeInvCompFields}.

Third, in the event that $G\supset H$, then one has to assume that the $H$-bundle $\P$ is a subbundle of a $G$-bundle, and \emph{mutatis mutandis} the proposition still holds. Such a situation occurs for instance when $\P$ is a reduction of a frame bundle  (of unspecified order), see Section~\ref{From tetrad to metric formulation of General Relativity} for an example.

Notice that despite the formal similarity with \eqref{ActiveGT} (or \eqref{PassiveGT}), the composite fields \eqref{CompFields} are not gauge transformed fields. Indeed the defining equivariance property \eqref{DF} of the dressing field implies $u\notin \H$, and $f\notin \Aut_v(\P)$. As a consequence, in general the composite fields do not belong to the gauge orbits of the original fields: $\chi^u \notin \mathcal{O}(\chi)$. Another consequence is that the dressing field method must also not be confused with a simple gauge fixing.

\subsection{Residual gauge symmetry}  
\label{Residual gauge symmetry}  

Suppose there is a normal subgroup $K$ and a subgroup $J$  of $H$ such that any $h \in H$ can be uniquely written as $h=jk$ for $j \in J$ and $k \in K$. Then $H=JK$ and $J \simeq H/K$, whose Lie algebra is denoted by $\LieJ$. Such a situation occurs for instance with $H = J \times K$. Several examples are based on this structure, see for instance Sections~\ref{The electroweak sector of the Standard Model} and \ref{Conformal Cartan geometry, tractors and twistors}.

The quotient bundle $\P/K$ is then a $J$-principal bundle $\P'=\P'(\M, J)$, with gauge group $\J \simeq \Aut_v(\P')$. The residual gauge symmetry of the composite fields depends on the one hand on that of the gauge fields, and on the other hand on that of the dressing field. A classification of the manifold of possible situations is impractical, but below we provide the general treatment of two most interesting cases. 
 
 \subsubsection{The composite fields as genuine gauge fields}  
 \label{The composite fields as genuine gauge fields}  

With the previous decomposition of $H$, our first case is summarized in this next result.
\begin{prop}  
\label{Prop2}
Let $u$ be a $K$-dressing field on $\P$. Suppose its $J$-equivariance  is given by 
\begin{align}
\label{CompCond}
R^*_ju=\Ad_{j\-}u, \quad \text{ for any } j \in J. 
\end{align}
Then the dressed connection $\omega^u$ is  a $J$-principal connection on $\P'$. That is, for $ X\in \LieJ$ and  $j\in J$, $\omega^u$ satisfies: $\omega^u(X^v)=X$ and $R^*_j\omega^u=\Ad_{j\-}\omega^u$. Its curvature is given by $\Omega^u$. 
Also, $\vphi^u$ is a $(\rho, V)$-tensorial map on $\P'$ and can be seen as a section of the associated bundle $E'=\P' \times_{J} V$. The covariant derivative on such sections is given by $D^u=d + \rho(\omega^u)$. 
\end{prop}

\noindent From this we immediately deduce the following

\begin{cor}
\label{Cor3}
The transformation of the composite fields under the residual $\J$-gauge symmetry is found in the usual way to be
\begin{align}
\label{GTCompFields}
(\omega^u)^{\gamma'} &:={\Phi'}^*\omega^u={\gamma'}\- \omega^u \gamma' + {\gamma'}\-d\gamma',  
&
(\vphi^u)^{\gamma'}&:={\Phi'}^*\vphi^u=\rho({\gamma'}\- )\vphi^u, \notag
\\
(\Omega^u)^{\gamma'} &:={\Phi'}^*\Omega^u={\gamma'}\- \Omega^u \gamma', 
 &
(D^u\vphi^u)^{\gamma'} &:={\Phi'}^*D^u\vphi^u=\rho({\gamma'}\-) D^u\vphi^u,
\end{align}
with $\Phi' \in \Aut(\P') \simeq \J \ni \gamma'$. 
\end{cor}

A quick way to convince oneself of this is to observe that for $\Phi' \in \Aut_v(\P')$ one has, using \eqref{CompCond},  $\big(u^{\gamma'}\big)(p):=({\Phi'}^*u)(p)=u(\Phi'(p))=u(p\gamma'(p))={\gamma'}(p)\-u(p)\gamma'(p)=({\gamma'}\- u \gamma')(p)$. 
 So, using again the generic variable $\chi$ one finds that $(\chi^u)^{\gamma'}=(\chi^{\gamma'})^{u^{\gamma'}}=(\chi^{\gamma'})^{{\gamma'}\-u \gamma'}=\chi^{u\gamma'}$, which proves \eqref{GTCompFields}. 
 In field theory, the relation $u^{\gamma'}={\gamma'}\- u \gamma'$ can be preferred  to \eqref{CompCond} as a  condition on the dressing field $u$.

The above results show that when \eqref{CompCond} holds, the composite fields \eqref{CompFields} are $\K$-invariant but genuine $\J$-gauge fields with residual gauge transformation given by \eqref{GTCompFields}. It may then be possible to perform a  further dressing operation provided  a suitable dressing field exists and satisfies the compatibility condition of being invariant under the $\K$-gauge subgroup just erased.  The extension of this scheme to any number of dressing fields can be found in \cite{Francois2014}. Let us now turn to our next interesting case.

 \subsubsection{The composite fields as twisted-gauge fields}  
 \label{The composite fields as twisted-gauge fields}  
 
 To define these gauge fields with a new behavior under the action of the gauge group, we need to introduce some definitions. Let $G' \supset G$ be a Lie group for which the representation $(\rho, V)$ of $G$ is also a representation of $G'$. Let us assume the existence of a $C^\infty$-map $C :\P \times J \rarrow G'$, $(p, j) \mapsto C_p(j)$, satisfying
\begin{align}
\label{PropDefC}
C_p(jj')=C_p(j)C_{pj}(j').
\end{align} 
From this we have that $C_p(e)=e$, with $e$ the identity in both $J$ and $G'$, and $C_p(j)\-=C_{pj}(j\-)$. Its differential is 
\begin{align*}
dC_{|(p, j)}=dC(j)_{|p} + dC_{p| j}: T_p\P \oplus T_jJ \rarrow T_{C_p(j)}G',
\end{align*}
where $\ker dC(j)=T_jJ$ and $\ker dC_p=T_p\P$, and where $dC(j)$ (resp. $dC_p$) uses the differential on $\P$ (resp. $J$). Notice that $C_p(j)\-dC_{|(p,j)} : T_p\P \oplus T_jJ \rarrow  T_eG'=\LieG'$. We then state the following result.
\begin{prop}
\label{P4}
Let $u$ be a $K$-dressing field on $\P$. Suppose its $J$-equivariance is given by
\begin{align}
\label{CompCond2}
(R_j^*u)(p)=j\-u(p)C_p(j),\quad \text{with $j\in J$ and  $C$ a map as above}. 
\end{align}
Then $\omega^u$ satisfies
\begin{enumerate}
\item $\omega^u_p(X^v_p)=c_p(X):=\tfrac{d}{dt}C_p(e^{tX})|_{t=0}$, $\quad$ for $X\in \LieJ$ and $X^v_p \in V_p\P'$.
\vspace{5pt}\item $R^*_j\omega^u=C(j)\- \omega^u C(j)+ C(j)\-dC(j)$.
\end{enumerate}
The dressed  curvature $\Omega^u$ is $J$-horizontal and satisfies $R^*_j\Omega^u=C(j)\-\Omega^uC(j)$. 
Also, $\vphi^u$ is a $\rho(C)$-equivariant map, $R^*_j \vphi^u=\rho\left(C(j)\right)\- \vphi^u$. The first order differential operator $D^u:=d + \rho_*(\omega^u)$ is a natural covariant derivative on such $\vphi^u$ so that $D^u\vphi^u$ is a $(\rho(C), V)$-tensorial form: 
$R^*_jD^u\vphi^u=\rho\left(C(j)\right)\- D^u\vphi^u$ and $(D^u\vphi^u)_p(X_p^v)=0$. 
\end{prop}
       
This proposition shows that $\omega^u$ behaves “almost as a connection”: we call it a $C$-twisted connection $1$-form. There is a natural geometric structure to interpret the dressed field $\vphi^u$. Omitting the representation $\rho$ of $G'$ on $V$ to simplify notations, we can define the following equivalence relation on $\P \times V$: 
\begin{equation*}
(p,v) \sim (pj, C_p(j)^{-1} v) \text{ for any $p\in\P$, $v \in V$, and $j \in J$.}
\end{equation*}
Using the properties of the map $C$, it is easy to show that this is indeed an equivalence relation. In particular, one has $(p j j', C_p(j j')^{-1} v) \sim(p j j',  C_{p j}(j')^{-1} C_p(j)^{-1} v) \sim (p j,  C_p(j)^{-1} v) \sim (p,v)$.
Then one can define the quotient vector bundle over $\M$
\begin{equation}
\label{C twisted vector bundle}
E = \P\times_{C(J)} V  := (\P \times V)/\mathord{\sim} 
\end{equation}
that we call a $C(J)$-twisted associated vector bundle to $\P$. Notice that when $J = \{e\}$, one has $E = \P \times V$. Adapting standard arguments in fiber bundle theory, one can show that sections of $E$ are $C(J)$-equivariant maps
\begin{equation*}
\varphi : \P \to V \text{ such that } \varphi(p j) = C_p(j)^{-1} \varphi(p) \text{ for any $p\in\P$, $j \in J$.}
\end{equation*}

The dressing field $\varphi^u$ is then a section of $E$ satisfying $\varphi^u(pk) = \varphi^u(p)$ for any $p\in \P$ and $k \in K$ by construction.

\smallskip
We can now deduce the transformations of the composite fields under the residual gauge group $\J$. 
 Consider $\Phi \in \Aut_v(\P')\simeq  \J \ni \gamma$, where $\gamma : \P \to J$ satisfies $\gamma(pk) = \gamma(p)$ and $\gamma(pj) = j\- \gamma(p) j$ for any $p \in \P$, $k \in K$ and $j \in J$, and define the map $C(\gamma):\P \rarrow G'$, $p \mapsto C_p\left(\gamma(p)\right)$, given by the compositions
\begin{align*}
\xymatrix@R=0ex@C=30pt{
 {\P} \ar[r]^-{\Delta} 
&
{ \P \times \P} \ar[r]^-{\id \times \gamma}
&
{\P \times J} \ar[r]^-{C}
&
{G'}
\\
 {p} \ar@{|->}[r] 
&
{(p,p)} \ar@{|->}[r]
&
{(p, \gamma(p))} \ar@{|->}[r]
&
{C_p(\gamma(p))}
 }
\end{align*}
Its differential  $dC(\gamma)_{|p} : T_p\P \rarrow T_{C_p\left(\gamma(p)\right)}G'$ is given by $dC(\gamma)=dC \circ \left( \id \oplus d\gamma \right) \circ d\Delta$, and we have $C_p\left(\gamma(p) \right)\-dC(\gamma)_{|p} : T_p\P \rarrow T_eG'=\LieG'$.  The residual gauge transformation of the dressing field is then $\big(u^{\gamma} \big)(p):=(\Phi^*u)(p)=u(p\gamma(p))=\gamma(p)\- u(p) C_p\big(\gamma(p)\big)=\big( {\gamma}\- u C(\gamma) \big) (p)$, that is
\begin{align}
\label{GT-CDressField}
u^{\gamma} = {\gamma}\- u C(\gamma).
\end{align}
This relation can be taken as an alternative to \eqref{CompCond2} as a  condition on the dressing field $u$. We  have then the following proposition.

\begin{prop} 
\label{P5}
Given  $\Phi \in \Aut_v(\P')\simeq  \J \ni \gamma$, the residual gauge transformations of the composite fields are
\begin{align}
\label{GTCompFields2}
(\omega^u)^{\gamma} &:= \Phi^*\omega^u = C(\gamma)\- \omega^u C(\gamma) + C(\gamma)\-dC(\gamma), 
\notag \\
(\vphi^u)^{\gamma}&:= \Phi^*\vphi^u = \rho\left( C(\gamma) \- \right) \vphi^u,
\notag \\
(\Omega^u)^{\gamma} &:=\Phi^*\Omega^u= C(\gamma) \- \Omega^u C(\gamma), 
\notag \\
 (D^u\vphi^u)^{\gamma} &:= \Phi^* D^u\vphi^u = \rho\left( C(\gamma) \- \right) D^u\vphi^u.
\end{align}
\end{prop}

This shows that the composite fields \eqref{CompFields} behave as \emph{gauge fields of a new kind}, on which the implementation of the \emph{gauge principle} is factorized through the map $C$. Given \eqref{GT-CDressField} and the usual $\J$-gauge transformations for the standard gauge fields $\chi$, the above results can be obtained by a direct algebraic calculation: $(\chi^u)^{\gamma}=(\chi^{\gamma})^{u^{\gamma}}=(\chi^{\gamma})^{{{\gamma}\-}uC(\gamma)}=\chi^{uC(\gamma)}$.

Under a further gauge transformation $\Psi \in \Aut_v(\P') \simeq \eta\in \J$, there are two ways to compute the composition $\Psi^*(\Phi^*u)$ of the two actions: first we use the composition inside the gauge group, $(\Phi \circ \Psi)(p) = p \gamma(p) \eta(p)$, so that $\left(\Psi^*(\Phi^*u)\right)(p) = \left( (\Phi \circ \Psi)^*u \right)(p) =u\left(p\gamma(p)\eta(p)\right) = \eta(p)\-\gamma(p)\- u(p)C_p\big(\gamma(p)\eta(p)\big)$; secondly, we compute the actions successively,
\begin{align*}
\left(\Psi^*(\Phi^*u)\right)(p)
&= \left( \gamma\- u C(\gamma) \right) (\Psi(p))
=\gamma\left(p\eta(p) \right)\- u\left(p\eta(p)\right) C_{p\eta(p)}\left(\gamma(p\eta(p) \right)
\\
&= \eta(p)\-\gamma(p)\-\eta(p) \cdot \eta(p)\-u(p) C_p\left( \eta(p)\right) \cdot C_{p\eta(p)}\left(\eta(p)\- \gamma(p) \eta(p) \right)
\\
&= \eta(p)\-\gamma(p)\- u(p)C_p\left(\gamma(p)\eta(p)\right).
\end{align*}
In both cases, $\Psi^*(\Phi^*u) = \eta\-\gamma\-\ u\ C\left(\gamma\eta \right)$, which secures the fact that the actions  \eqref{GTCompFields2} of the residual gauge symmetry on the composite fields are well behaved as representations of the residual gauge group, \emph{even if $C$ is not a morphism of groups}.

Ordinary connections correspond to $C_p(j) = j$ for any $p \in \P'$ and $j \in J$, in which case, it is a morphism of groups.

\paragraph{The case of  $1$-$\alpha$-cocycles.} For a $p \in \P'$, suppose given $C_p : J \rarrow G'$ satisfying $C_p(jj')=C_p(j)\ \alpha_j[C_p(j')]$ for $\alpha: J \rarrow \Aut(G')$ a continuous group morphism. Such an object appears in the representation theory of crossed products of $C^*$-algebras and is known as a \emph{$1$-$\alpha$-cocycle} (see \cite{Pedersen79,Williams07}).\footnote{In the general theory the group $G'$ is replaced by a $C^*$-algebra $A$.} Then, defining $C_{pj}(j') := \alpha_j[C_p(j')]$, one has an example of \eqref{PropDefC}, and the above result applies to the $1$-$\alpha$-cocycle $C$. As a particular case, consider the following
\begin{prop}
\label{alpha_cocycle}
 Suppose $J$ is abelian and let $A_p, B : J\rarrow GL_n$  be group morphisms where $R^*_jA_p(j')=B(j)\- A_p(j') B(j)$. Then $C_p:=A_pB : J\rarrow GL_n$ is a $1$-$\alpha$-cocyle with $\alpha: J \rarrow \Aut(GL_n)$ defined by $\alpha_j [g] =B(j)\- g B(j)$ for any $g \in GL_n$.
 \end{prop}

\noindent Using the commutativity of $J$ through $B(j)B(j') = B(jj') = B(j'j) = B(j')B(j)$, the proposition is proven as $C_p(jj') =A_p(jj')B(jj')=A_p(j)A_p(j')B(j)B(j') =A_p(j)B(j)\ B(j)\-[ A_p(j')B(j')] B(j)=C_p(j)\ B(j)\-[C_p(j')]B(j)$. Notice also that we have $C_p(jj')=C_p(j'j)=C_p(j')\ B(j')\-[C_p(j)]B(j')$.  Such $1$-$\alpha$-cocycles will appear in the case of the conformal Cartan geometry and the associated Tractors and Twistors in Section~\ref{Conformal Cartan geometry, tractors and twistors}.

\subsection{Application to the BRST framework} 
\label{Application to the BRST framework} 

\subsubsection{The BRST differential algebra}  
\label{The BRST differential algebra}  

The  BRST differential algebra captures the infinitesimal version of \eqref{ActiveGT}. Abstractly (see for instance \cite{Dubois-Violette1987}) it is a   bigraded  differential algebra generated by  $\{\omega, \Omega, v, \zeta\}$  where $v$ is the so-called ghost and the generators are respectively of degrees $(1, 0)$, $(2, 0)$, $(0, 1)$ and $(1, 1)$.  It is endowed with two nilpotent antiderivations $d$ and $s$, homogeneous of degrees $(1, 0)$ and $(0, 1)$ respectively, with vanishing anticommutator: $d^2=0=s^2$, $sd+ds=0$.  The algebra is equipped with a bigraded commutator $[\alpha, \beta]:=\alpha\beta -(-)^{\text{deg}[\alpha]\text{deg}[\beta]}\beta\alpha$. The action of $d$ is defined on the generators by: $d\omega=\Omega -\tfrac{1}{2}[\omega, \omega]$ (Cartan structure equation), $d\Omega=[\Omega, \omega]$ (Bianchi identity), $dv=\zeta$ and $d\zeta=0$. The action of the BRST operator on the generators gives the usual defining relations of the BRST algebra,
\begin{align}
\label{BRST}
s\omega=-dv -[\omega, v], \quad s\Omega=[\Omega, v], \quad  \text{ and } \quad sv=-\tfrac{1}{2}[v, v]. 
\end{align}

When the abstract BRST algebra is realized in a differential geometry framework, the bigrading is according to the de Rham form degree and  ghost degree: $d$ is the de Rham differential on $\P$ (or $\M$ if one works in local trivializations) and $s$ is the de Rham differential on $\H$. The ghost is the Maurer-Cartan form on $\H$ so that $v \in \bigwedge^1(\H, \text{Lie}\H)$, and given $\xi \in T\H$, $v(\xi) :\P \rarrow \LieH  \in \text{Lie}\H$ \cite{Bonora-Cotta-Ramusino}.  So in practice the ghost can be seen as a map $v:  \P\rarrow \LieH \in \text{Lie}\H$, a placeholder that takes over the role of the infinitesimal gauge parameter. Thus the first two  relations of \eqref{BRST} (and \eqref{BRST2} below) reproduce the infinitesimal gauge transformations of the gauge fields \eqref{ActiveGT}, while the third equation in \eqref{BRST} is the Maurer-Cartan structure equation for the gauge group $\H$. The BRST transformation of the section $\vphi$ (of degrees $(0, 0)$) and its covariant derivative are
\begin{align}
\label{BRST2}
s\vphi=-\rho_*(v)\vphi, \quad \text{ and } \quad sD\vphi=-\rho_*(v)D\vphi.
\end{align}
where $\rho_*$ is the representation of the Lie algebra induced by the representation $\rho$ of the group.

The BRST framework provides an algebraic characterization of relevant quantities in gauge theories, such as admissible Lagrangian forms, observables and anomalies, all of which required to belong to the $s$-cohomology group $H^{*, *}(s)$ of $s$-closed but not $s$-exact quantities.

\subsubsection{Modified BRST differential algebra}  
\label{Modified BRST differential algebra}  

Since the BRST algebra encodes the infinitesimal gauge transformations of the gauge fields, it is expected that the dressing field method modifies it. To see how, let us first consider the following 
\begin{prop}    
Consider the BRST algebra \eqref{BRST}-\eqref{BRST2} on the initial gauge variables and the ghost $v\in Lie\H$. Introducing the \emph{dressed ghost}
\begin{equation}
\label{NewBRST}
v^u=u\-vu + u\-su,
\end{equation}
the composite fields \eqref{CompFields} satisfy the  \emph{modified BRST algebra}:
\begin{align*}
s\omega^u &=-D^uv^u=-dv^u-[\omega^u, v^u], 
&
s\vphi^u &=-\rho_*(v^u)\vphi^u,
\notag\\
s\Omega^u &=[\Omega^u, v^u], 
&
sv^u &=-\tfrac{1}{2}[v^u, v^u].
\end{align*}
This result does not rest on the assumption that $u$ is a dressing field. 
\end{prop}

The result is easily found by expressing the initial gauge variable $\chi=\{ \omega, \Omega, \vphi\}$ in terms of the dressed fields $\chi^u$ and the dressing field $u$, and re-injecting in the initial BRST algebra  \eqref{BRST}-\eqref{BRST2}. At no point of the derivation does $su$ need to be explicitly known. It then holds regardless if $u$ is a dressing field or not. 

If the ghost $v$ encodes the infinitesimal initial $\H$-gauge symmetry, the dressed ghost $v^u$ encodes the infinitesimal residual gauge symmetry. Its concrete expression depends on the BRST transformation of $u$. 

Under the hypothesis $K\subset H$, the ghost decomposes as $v=v_\LieK + v_{\LieH/\LieK}$, and the BRST operator splits accordingly: $s=s_\LieK + s_{\LieH/\LieK}$. If $u$ is a dressing field its BRST transformation is the infinitesimal version of its defining transformation property: $s_\LieK u=-v_\LieK u$. So the dressed ghost is
\begin{align*}
v^u &=u\-vu+u\-su=u\-(v_\LieK + v_{\LieH/\LieK})u+u\-(-v_\LieK u+ s_{\LieH/\LieK}u)
\\
&=u\- v_{\LieH/\LieK} u+u\- s_{\LieH/\LieK}u. 
\end{align*}
The Lie$\K$ part of the ghost, $v_\LieK$, has disappeared. This means that $s_\LieK \chi^u=0$, which expresses the $\K$-invariance of the composite fields \eqref{CompFields}. 

\paragraph{Residual BRST symmetry}  

If $K \subset H$ is a normal subgroup, then $H/K=J$ is a group with Lie algebra $\LieH/\LieK=\LieJ$. We here provide the BRST treatment of the two cases detailed in Section~\ref{Residual gauge symmetry}.

Suppose the dressing field satisfies the condition \eqref{CompCond}, whose BRST version is $s_{\LieJ} u=[u, v_{\LieJ}]$. The dressed ghost is then
\begin{align}
\label{NewGhost2}
v^u=u\- v_{\LieJ} u+u\- s_{\LieJ}u=u\- v_{\LieJ} u+u\- (u v_{\LieJ} -  v_{\LieJ}u)=v_{\LieJ}. 
\end{align}
This in turn implies that the new BRST algebra is
\begin{align}
\label{NewBRST2}
s\omega^u &=-D^uv_{\LieJ}=-dv_{\LieJ}-[\omega^u, v_{\LieJ}], 
&
s\vphi^u &=-\rho_*(v_{\LieJ})\vphi^u,
\notag\\
s\Omega^u &=[\Omega^u, v_{\LieJ}], 
&
sv_{\LieJ} &=-\tfrac{1}{2}[v_{\LieJ}, v_{\LieJ}].
\end{align}
This is the BRST version of \eqref{GTCompFields}, and reflects the fact that the composite fields \eqref{CompFields} are genuine $\J$-gauge fields, in particular that $\omega^u$ is a $J$-connection. 

Suppose now that the dressing field satisfies the  condition \eqref{CompCond2}, whose BRST version is $s_{\LieJ} u=-v_{\LieJ}u + uc_p(v_{\LieJ})$. The dressed ghost is then
\begin{align}
\label{NewGhost3}
v^u=u\- v_{\LieJ} u+u\- s_{\LieJ}u=u\- v_{\LieJ} u+u\- \left(-v_{\LieJ}u + uc_p(v_{\LieJ})\right)=c_p(v_{\LieJ}). 
\end{align}
This in turn implies that the new BRST algebra is
\begin{align}
\label{NewBRST3}
s\omega^u &=-dc_p(v_{\LieJ})-[\omega^u, c_p(v_{\LieJ})], 
&
s\vphi^u &=-\rho_*(c_p(v_{\LieJ}))\vphi^u,
\notag\\
s\Omega^u &=[\Omega^u, c_p(v_{\LieJ})], 
&
sc_p(v_{\LieJ}) &=-\tfrac{1}{2}[c_p(v_{\LieJ}), c_p(v_{\LieJ})]. 
\end{align}
This is the BRST version of \eqref{GTCompFields2}, and reflects the fact that the composite fields \eqref{CompFields} instantiate the gauge principle in a satisfactory way.

To conclude we mention that the dressing operation is compatible with Stora's method of altering a BRST algebra so that it describes the action of infinitesimal diffeomorphisms of the base manifold on the gauge fields, in addition to their gauge transformations, as described in \cite{Langouche1984, Stora:2005tp} for instance: details can be found in \cite{FLM2016_I}.

\subsection{Local construction and physics} 
\label{Local construction and physics} 

 Until now, we have been focused on the global aspects of the dressing approach on the bundle $\P$  to emphasize the geometric nature of the composite fields obtained. Most notably we showed that  the composite field can behave as “generalized” gauge fields.  But to do physics we need the local representatives on an open subset $\U\subset \M$ of global dressing and composite fields. These are obtained in the usual way from a local section $\sigma:\U \rarrow \P$ of the bundle. The  important properties they thus retain are their gauge invariance and their residual gauge transformations. 

If it happens that a dressing field is defined locally on $\U$ first, and not directly on $\P$, then the local composite fields $\chi^u$ are defined in terms of the local dressing field $u$ and local gauge fields $\chi$ by \eqref{CompFields}. The gauge invariance and residual gauge transformations of these local composite fields are derived from the gauge transformations of the local dressing field under the various subgroups of the local gauge group $\H_\text{\tiny{loc}}$ according to $(\chi^u)^\gamma=(\chi^\gamma)^{u^\gamma}$. The BRST treatment for the local objects mirrors exactly the one given for global objects.  

This being said, note $A=\sigma^*\omega$ and $F=\sigma^*\Omega$ for definiteness but keep $u$ and $\vphi$ to denote the local dressing field and sections of the associated vector bundle $E$. Suppose that the base manifold is equipped with a $(r, s)$-Lorentzian metric allowing for a Hodge star operator, and that $V$ is equipped with an inner product $\langle\ , \rangle$. We state the final proposition dealing with gauge theory.

\begin{prop} 
\label{Prop-Lagrangian}
Given the geometry defined by a bundle $\P(\M, H)$ endowed with $\omega$ and the associated bundle $E$, suppose we have a gauge theory given by the prototypical $\H_\text{\tiny{loc}}$-invariant Yang-Mills Lagrangian
\begin{align*}
L(A, \vphi)=\tfrac{1}{2}\Tr(F \w * F) + \langle D\vphi, \ *D\vphi\rangle - U(\norm{\vphi}) \vol, 
\end{align*}
where $\vol$ is the metric volume form on $\M$, $\norm{\vphi}:=|\langle \vphi \rangle |^{\sfrac{1}{2}}$ and $U$ is a potential term.\footnote{For instance, such a term is the one for a spontaneous symmetry breaking mechanism.}. If there is a local dressing field $u: \U \rarrow G  \subset H$ with $\K_\text{\tiny{loc}}$-gauge transformation $u^\gamma=\gamma\-u$, then the above Lagrangian is actually  a “$\H_\text{\tiny{loc}}/\K_\text{\tiny{loc}}$-gauge theory” defined in terms of  $\K_\text{\tiny{loc}}$-invariant variables since we have
\begin{align*}
L(A, \vphi)=L(A^u, \vphi^u)=\tfrac{1}{2}\Tr(F^u \w * F^u) + \langle D^u\vphi^u, \ *D^u\vphi^u\rangle - U(\norm{\vphi^u}) \vol.
\end{align*}
\end{prop}

The relation $L(A, \vphi)=L(A^u, \vphi^u)$ is satisfied since, as already noticed, relations \eqref{CompFields} look algebraically like gauge transformations \eqref{ActiveGT} under which $L$ is supposed to be invariant in a formal way.

The terminology “$\H_\text{\tiny{loc}}/\K_\text{\tiny{loc}}$-gauge theory” means that the Lagrangian is written in terms of fields which are invariant under the action of $\gamma:\U \rarrow K$. Since the quotient $H/K$ needs not be a group, the remaining symmetries of the fields might not be described in terms of a group action.

Notice  that since $u$ is a dressing field, $u \notin \H_\text{\tiny{loc}}$, so the dressed Lagrangian $L(A^u, \vphi^u)$ ought not to be confused with a gauge-fixed Lagrangian $L(A^\gamma, \vphi^\gamma)$ for some chosen $\gamma \in \H_\text{\tiny{loc}}$, even if it may happen that $\gamma=u$ as fields if one forgets about the corresponding representations of the gauge group, a fact that might go unnoticed. As we have stressed in Section~\ref{Reduction of gauge symmetries: the dressing field method}, the dressing field approach is distinct from both gauge-fixing and spontaneous symmetry breaking as a means to reduce gauge symmetries. 

 Let us highlight the fact that a dressing field can often be constructed  by requiring the gauge invariance of a prescribed ``gauge-like condition''. Such a condition is given when a local gauge field $\chi$ (often the gauge potential)  transformed by a field $u$ with value in the symmetry group $H$, or one of its subgroups, is required to satisfy a functional constraint: $\Sigma(\chi^u)=0$.  Explicitly solved, this makes $u$ a function of $\chi$, $u(\chi)$, thus sometimes called \emph{field dependent gauge transformation}. However this terminology is valid if and only if  $u(\chi)$ transforms under the action of $\gamma\in \H_\text{\tiny{loc}}$ as $u(\chi)^\gamma:=u(\chi^\gamma)=\gamma\- u(\chi) \gamma$, in which case $u(\chi) \in \H_\text{\tiny{loc}}$.  But if the functional constraint  still holds under the action of $\H_\text{\tiny{loc}}$, or of a subgroup thereof, it follows  that $(\chi^\gamma)^{u^\gamma}=\chi^u$ (or equivalently that $s\chi^u=0$). This in turn suggests that $u^\gamma=\gamma\- u$ (or $su=-vu$) so that $u \notin \H_\text{\tiny{loc}}$ but is indeed a dressing field.
  
This, and the above proposition, generalizes the pioneering idea of Dirac \cite{Dirac55, Dirac58}  aiming at quantizing QED by rewriting the classical theory in terms of gauge-invariant variables. The idea was rediscovered several times and sometimes termed \emph{Dirac variables} \cite{Pervushin, Lantsman}. They reappeared in various contexts in gauge theory, such as  QED \cite{Lavelle-McMullan93},  quarks theory in QCD \cite{McMullan-Lavelle97}, the proton spin decomposition controversy  \cite{LorceGeomApproach, Leader-Lorce, FLM2015_I}. The dressing field approach thus gives a unifying and clarifying framework for these works, and others concerning the BRST treatment of anomalies in QFT \cite{Manes-Stora-Zumino1985, Garajeu-Grimm-Lazzarini}, Polyakov's ``partial gauge fixing'' for $2D$-quantum gravity \cite{Polyakov1989, Lazzarini2008} or the construction of the Wezz-Zumino functionnal \cite{Attard-Lazz2016}.

In the following we provide examples of significant applications of the dressing field approach in various contexts: the electroweak sector of the Standard Model, the tetrad \textit{vs} metric formulation of GR, and  tractors and twistors obtained from conformal Cartan geometry.

\section{The electroweak sector of the Standard Model} 
\label{The electroweak sector of the Standard Model} 

The aim of the electroweak  model is to give a gauge theoretic account of the fact that there is one long range interaction mediated by a massless boson, electromagnetism, together with a short range interaction mediated by massive bosons, the weak interaction. Here we discard the spinors (matter fields) of the theory and consider only the theory describing the gauge potentials and the scalar field. The spinors could be treated along the lines of the following exposition. More details can be found in \cite{Masson-Wallet, Francois2014}. 

\subsection{Reduction of the $\SU(2)$-symmetry via dressing}
\label{Reduction of the SU(2)-symmetry via dressing}

The principal bundle of the model is $\P\left(\M, U(1)\times SU(2)\right)$ and it is endowed with a connection whose local representative is $A=a+b$. Its curvature is $F=f_a+g_b$. The defining representation of the structure group is $(\CC^2, \ell)$, with $\ell$ the left matrix multiplication. The associated vector bundle is $E=\P\times_\ell \CC^2$ and we denote by $\vphi : \U\subset \M \rarrow \CC^2$ a (local) section. The covariant derivative is $D\vphi= d\vphi +(g'a +gb)\vphi$, with $g', g$ the coupling constants of $U(1)$ and $SU(2)$ respectively. The action of the gauge group $\H=\U(1) \times \SU(2)$ (we drop the subscript ``loc'' from now on) is,
\begin{align*}
a^\alpha &=a +\tfrac{1}{g'}\alpha^{-1}d\alpha,
&
b^\alpha &=b, 
&
\vphi^\alpha &=\alpha^{-1}\vphi,
\\
a^\beta &=a, 
&
b^\beta &=\beta^{-1}b\beta + \tfrac{1}{g}\beta\-d\beta,
&
\vphi^\beta&=\beta^{-1}\vphi,
\end{align*}
where $\alpha \in \U(1)$ and $\beta \in \SU(2)$. The structure of direct product group is clear. The $\H$-invariant Lagrangian form of the theory is, 
\begin{align}
\label{EW-Lagrangian}
L(a, b, \vphi)&=\tfrac{1}{2}\Tr(F \w * F) + \langle D\vphi, \ *D\vphi\rangle - U(\norm{\vphi}) \vol, \notag\\
			&=
			\begin{multlined}[t]
			\tfrac{1}{2}\Tr(f_a\wedge *f_a) +\tfrac{1}{2}\Tr(g_b\wedge *g_b) 
			\\[2pt]
			+ \langle D\vphi, \ *D\vphi\rangle -\left( \mu^2 \langle\vphi, \vphi\rangle +\lambda \langle\vphi, \vphi\rangle^2\right) \vol,
			\end{multlined}
\end{align}
where $\mu, \lambda \in \RR$. This gauge theory describes the interaction of a doublet scalar field $\vphi$ with two gauge potentials $a$ and $b$. As it stands, nor $a$ nor $b$ can be massive, and indeed $L$ contains no mass term for them. It is not a problem for $a$ since we expect to have at least one massless field to carry the electromagnetic interaction. But the weak interaction is short range, so its associated field must be massive. Hence the necessity to reduce the $SU(2)$ gauge symmetry in the theory in order to allow a mass term for the weak field. Of course we know that this can be achieve via SSBM. Actually the latter is used in conjunction with a gauge fixing, the so-called \textit{unitary gauge}, see e.g \cite{Becchi-Ridolfi}. Some authors have given a more geometrical account of the mechanism based on the bundle reductions theorem, see  \cite{Trautman, Westenholz, Sternberg}. 

We now show that the $\SU(2)$ symmetry can be erased via the dressing field method. Given the gauge transformations above, we define a dressing field out of the doublet scalar field $\vphi$ by using a polar decomposition $\vphi =u \eta$ in $\CC^2$ with
\begin{align}
\label{decomp-phi}
u\in SU(2) \quad \text{ and }\quad \eta:=  \vect{0 \\ \norm{\vphi}} \in \RR^+ \subset \CC^2, \quad \text{so that} \quad   u^\beta=\beta\- u,
\end{align}
as can be checked explicitly. Then  $u$ is a $\SU(2)$-dressing field that can be used to apply Prop.~\ref{P1} and to construct the $\SU(2)$-invariant composite fields
\begin{align}
\h A&=u\-Au+\tfrac{1}{g}u\-du= a+  (u\-bu +\tfrac{1}{g}u\-du) =: a + B,     \notag \\
\h F&=u\- F u=f_a+u\-g_bu =: f_a+G,\qquad \text{with $G=dB+gB^2$,}     \notag\\[1mm]
\h \vphi&=u\-\vphi=\eta, \qquad \text{and} \qquad \h D\h\vphi=u\-D\vphi=\h D\eta=d\eta + (g'a + gB)\eta.
\end{align}
By virtue of Prop.~\ref{Prop-Lagrangian}, we conclude that the theory defined by the electroweak Lagrangian \eqref{EW-Lagrangian} is actually a $\U(1)$-gauge theory described in terms of the above composite fields,
\begin{align}
\label{EW-Lagrangian2}
L(a, B, \eta)&=\tfrac{1}{2}\Tr(\h F \w * \h F) + \langle \h D\eta, \ *\h D\eta \rangle - U(\eta) \vol, \notag\\
			&=\tfrac{1}{2}\Tr(f_a\wedge *f_a) +\tfrac{1}{2}\Tr(G\wedge *G) + \langle \h D\eta, \ *\h D\eta \rangle -\left( \mu^2\eta^2 +\lambda \eta^4  \right) \vol.
\end{align}
Notice that by its very definition $\eta^\beta=\eta^\alpha=\eta$, so it is already a fully gauge invariant scalar field which  then qualifies as an observable. 

\subsection{Residual $\U(1)$-symmetry}
\label{Residual U(1)-symmetry}

Is a mass term allowed for the  $\SU(2)$-invariant field $B$? To answer one needs to check its  $\U(1)$-residual gauge transformation $B^\alpha$, which depends on the $\U(1)$-gauge transformation of the dressing field $u$. One can  check that
\begin{align*}
u^\alpha= u\t\alpha, \qquad \text{where} \qquad  \t\alpha= \begin{pmatrix} \alpha & 0  \\ 0 & \alpha\- \end{pmatrix}.
\end{align*}
We therefore have
\begin{align*}
B^\alpha&=(b^\alpha)^{u^\alpha}=\t\alpha\-u\- b u\t\alpha +\tfrac{1}{g}\t\alpha\-(u\-du)\t\alpha +\tfrac{1}{g}\t\alpha\-d\t\alpha
=\t\alpha\- B \t\alpha + \tfrac{1}{g}\t\alpha\-d\t\alpha, \\
G^\alpha&=(g_b^\alpha)^{u^\alpha}=\t\alpha\-u\-g_bu\t\alpha=\t\alpha\- G \t\alpha.
\end{align*}
In view of this, it would seem that $B$ still cannot have mass terms. But given the decomposition $B=B_a \sigma^a$ where $\sigma^a$ are the hermitian Pauli matrices and $B_a \in i\RR$, so that $\bar B_a =-B_a$, we have explicitly
\begin{align*}
B &=B_a\sigma^a=
\setlength\arraycolsep{5pt}
\begin{pmatrix} B_3 & B_1-iB_2 \\ B_1+iB_2 & -B_3  \end{pmatrix}
=:\begin{pmatrix} B_3 & W^- \\ W^+ & -B_3  \end{pmatrix}, 
\end{align*}
and
\begin{align*}
B^\alpha &=
\setlength\arraycolsep{5pt}
\begin{pmatrix}  B_3 +\frac{1}{g}\alpha\-d\alpha & \alpha^{-2}W^- \\[5pt] \alpha^2W^+ & -B_3 -\frac{1}{g}\alpha\-d\alpha \end{pmatrix}.
\end{align*}
The fields $W^\pm$ transform tensorially under $\U(1)$, and so they can be massive: they are the ($U(1)$-charged) particles detected in the SPS collider in January $1983$. The field $B_3$ transforms as a $U(1)$-connection, making it another massless field together with the genuine $U(1)$-connection $a$. Considering $(a, B_3)$ as a doublet, one can perform a natural change of variables
\begin{align*}
\vect{A \\ Z^0} 
\setlength\arraycolsep{5pt}
:=\begin{pmatrix} \cos\theta_W &  \sin\theta_W \\ -\sin\theta_W & \cos\theta_W \end{pmatrix} \vect{a \\ B_3} = \vect{\cos\theta_W a + \sin\theta_W B_3\\  \cos\theta_W B_3 - \sin\theta_W a},
\end{align*}
where the so-called Weinberg (or weak mixing) angle is defined by $\cos \theta_W = \sfrac{g}{\sqrt{g^2+g'^2}}$ and $\sin\theta_W=\sfrac{g'}{\sqrt{g^2+g'^2}}$. By construction, it is easy to show that the $1$-form $Z^0$  is then fully gauge invariant and can therefore be both massive  and observable: it is the neutral weak field whose boson has been detected in the SPS collider in May $1983$. Now, still by construction, we have $A^\beta=A$ and  $A^\alpha=A + \tfrac{1}{e} \alpha\-d\alpha$ with coupling constant $e :=\sfrac{gg'}{\sqrt{g^2+g'^2}}=g'\cos\theta_W=g\sin\theta_W$. So $A$ is a $U(1)$-connection: it is the massless carrier of the electromagnetic interaction and $e$ is the elementary electric charge. 

The electroweak theory \eqref{EW-Lagrangian2} is then expressed in terms of the gauge invariant fields $\eta, Z^0$ and of the $U(1)$-gauge fields $W^\pm, A$:
\begin{align}
\label{EW-Lagrangian3}
L(A, W^\pm, Z^0, \eta) &=
\tfrac{1}{2}\Tr(\h F \w * \h F) + \langle D\eta, \ *D\eta \rangle - U(\eta) \vol 
\notag\\[2mm]
&=
	dZ^0\wedge*dZ^0 + dA\wedge*dA + dW^-\wedge*dW^+  
\notag\\  
&\phantom{==}	+ 2g\bigg\{ 
	\begin{aligned}[t]		
      &\sin\theta_W \big(dA\wedge *(W^-W^+)
      +  \cos\theta_W\big( dZ^0\wedge*(W^-W^+) 
     \\
      &+dW^-\wedge*(W^+A)
       +dW^-\wedge*(W^+Z^0) 
       \\
     &+dW^+\wedge*(AW^-) \big)  
      +dW^+\wedge*(Z^0W^-) \big) 
      \bigg\} 
	\end{aligned}
\notag\\
 &\phantom{==}     +4g^2\bigg\{ 
	\begin{aligned}[t]		
      &\sin^2\theta_W\  AW^-\wedge*(W^+A) 
      \\
      &+ \cos^2\theta_W\  Z^0W^-   \wedge*(W^+Z^0) 
	\\
	&+ \sin\theta_W\cos\theta_W\  AW^-\wedge*(W^+Z^0) 
	\\
      	&+ \sin\theta_W\cos\theta_W\  Z^0W^-\wedge*(W^+A)   
	\\
	&+  \frac{1}{4}\ W^-W^+\wedge*(W^-W^+)    
      \bigg\}   
	\end{aligned}
\notag\\
&\phantom{==}    + d\eta\wedge*d\eta  
    	-  g^2\eta^2\  W^+\wedge*W^-    
	-  (g^2+g'^2)\eta^2\  Z^0\wedge*Z^0
\notag\\
&\phantom{==}  -\left( \mu^2\eta^2 +\lambda \eta^4  \right) \vol.
\end{align}
We can read off all possible interactions between the four electroweak fields. Notice that there is no coupling between the fields $A$ and $Z^0$, showing the electric neutrality of the $Z^0$.

The next natural step is to expand the $\RR^+$-valued scalar field $\eta$ around its \emph{unique} configuration $\eta_0$ minimizing the  potential $U(\eta)$, the so-called Vacuum Expectation Value (VEV), as $\eta=\eta_0+H$ where $H$ is the gauge invariant Higgs field. True mass terms for  $Z^0, W^\pm$ and $H$ depending on $\eta_0$ then appear from the couplings of the electroweak fields with $\eta$ and from the latter's self interaction. The absence of coupling between $\eta$ and $A$ indicates the masslessness of the latter (the two photons decay channel of the Higgs boson involves intermediary leptons, not treated here).  

The theory has two qualitatively distinct phases. In the phase  where $\mu^2>0$, the VEV vanishes and  so  do all masses, while in the phase where $\mu^2<0$, the VEV is non-vanishing, $\eta_0=\sqrt{-\mu^2/2\lambda}$.  The masses of the fields $Z^0, W^\pm$  and $H$ are then $m_{Z_0}=\eta_0\sqrt{(g^2+g'^2)}$,  $m_{W^\pm}=\eta_0g$ ,  with ratio $\tfrac{m_{W^\pm}}{m_{Z^0}}=\cos\theta_W$, and $m_H=\eta_02\lambda$. In this case,  \eqref{EW-Lagrangian3} becomes the electroweak Lagrangian form of the Standard Model in the so-called unitary gauge. But keep in mind that, as a result of the dressing field method, no gauge fixing nor SSBM is involved to obtain it.

\subsection{Discussion}
\label{Discussion}

Some differences with the usual viewpoint is worth stressing. The SSBM is usually constructed as follows. At high energy (\textit{i.e.} in the phase $\mu^2>0$) the symmetric VEV $\vphi_0=\binom{0}{0}$ of $\vphi\in \CC^2$ respect the full $\SU(2)\times \U(1)$ gauge symmetry group so that no gauge potential in the theory can be massive. At low energy (\textit{i.e.} in the phase $\mu^2<0$) the field $\vphi$ must fall somewhere in the space of configurations that minimize the potential $U(\vphi)$. A space which is a circle in $\CC^2$ defined by $M_0=\big\{\vphi\in \CC^2\ | \ \bar\vphi_1\vphi_1 +\bar \vphi_2\vphi_2=-\mu^2/\lambda\big\}$, whose individual points are not invariant under $\SU(2)$. Then, once an arbitrary minimum $\vphi_0\in M_0$ is \emph{randomly} selected, the gauge group is broken down to $\U(1)$ and mass terms for $SU(2)$-gauge potentials are generated. See e.g \cite{Zinn-Justin}. This usual interpretation takes place in the \emph{history} of the Universe, and this “phase transition” is a contingent phenomena, since it selects by chance one specific value in $M_0$. The Standard Model of Particles Physics (SMPP) then relies on two strong foundations: one is structural (in the mathematical way), it is the Lagrangian of the theory; the other one is contingent, it is the historical aspect of the SSBM. 
 
The dressing field approach allows to clearly distinguish the \emph{erasure of $\SU(2)$} and the \emph{generation of  mass terms} as two distinct operations, the former being a prerequisite of the latter  but not its direct \textit{cause}, as the textbook interpretation would have it. Notice also that the relevant $\SU(2)$-invariant variables, corresponding to the physical fields (fermions fields are treated in the same manner, see \cite{Masson-Wallet}), are identified at the mathematical level of the theory in both phases (\textit{i.e.} independently of the sign of $\mu^2$).  The transition between these phases, characterized by different electroweak vacuum, remains a dynamical process parametrized  by the sign of $\mu^2$.\footnote{In fact, it could even be reduced to a technical step useful to perform the usual field quantization procedure, which relies heavily on the identification of propagators and mass terms in the Lagrangian.} But we stress that in our scheme there is no arbitrariness in the choice of  VEV for $\eta \in \RR^+$ since it is now unique: $\eta_0 = \sqrt{-\mu^2/2 \lambda}$ when $\mu^2<0$. In particular, all the bosons $Z^0, A, W^+, W^-$ (and fermions fields) can be identified at the level of the theory, without requiring any historical contingent process. In that respect, the contingent aspect of the SMPP is dispelled to the benefit of its unique structural foundation.

The arbitrariness of the polar decomposition $\vphi =u \eta$ is discussed in \cite{Masson-Wallet}: imposing that the final $U(1)$ charges are clearly identified, the field content of the Lagrangian in the new variables is the same up to global transformations involving some rigid transformations of the fields. This implies that the content of the theory in terms of $\SU(2)$-invariant fields takes place at an ontological level, since it does not require any historical arguments.
 
According to \cite{Westenholz}, the very meaning of the terminology ``\textit{spontaneous} symmetry breaking'' lies in the fact that $M_0$ is not reduced to a point. Granting this reasonable observation, the dressing field approach would then lead to deny the soundness of this terminology to characterize the electroweak model. First because the symmetry reduction is not related to the choice of a VEV in $M_0$, then because the latter is reduced to a point. A better characterization would emphasize the link between mass generation and electroweak vacuum phase transition: ``mass generation through electroweak vacuum phase transition''.

\smallskip
The fact that the dressing field approach to the electroweak model allows to dispense with the idea of spontaneous breaking of a gauge symmetry is perfectly in line with the so-called Elitzur theorem stating that in lattice gauge theory a gauge symmetry cannot be spontaneously broken. An equivalent theorem for gauge field theory has not been proven, but no reason has been given as to why it would fail either. 

Furthermore, as we have mentioned in the introduction, the status of gauge symmetries is a disputed question in philosophy of physics. A well argued position considers gauge symmetries as ``surplus structures'', as philosopher of physics Michael Redhead calls it, that is a redundancy in our mathematical description of reality. They would then have an epistemological status. The idea of a spontaneous breakdown of  a gauge symmetry on the other hand, insofar as it implies observable qualitative physical effects (particles acquire masses in a historical process), supports an ontological view of gauge symmetries, making them a structural feature of reality rather than of our description of it. And indeed the part of the philosophy of physics community interested in this problem has struggled to reconcile the empirical success of the electroweak model with their analysis of gauge symmetries (see e.g  \cite{StanEncyPhil_Sym-SymBreak_Castellani_Brading, Castellani-Brading}). Often a workaround if proposed in arguing that a gauge fixing removes the local dependence of the symmetry and that only a global one remains to be broken spontaneously, which by the Goldstone theorem generates Goldstone bosons ``eaten up'' by the gauge bosons gaining masses in the process.  

These efforts of interpretation are enlightened once it is recognized that the notion of spontaneous breaking of gauge symmetry is not pivotal to the empirical success of the electroweak model. Higgs had a glimpse of this fact \cite{Higgs66}, and Kibble saw it clearly \cite{Kibble67} (see the paragraph just before the conclusion of his paper). Both had insights by working on toy models, just before the electroweak model was proposed by Weinberg and Salam in $1967$. The invariant version of the model was first given in \cite{Frohlich-Morchio-Strocchi81} in $1981$ (compare Section~6 with our exposition above), but was rediscovered independently by others \cite{McMullan-Lavelle95, Chernodub2008, Faddeev2009, Masson-Wallet, Ilderton-Lavelle-McMullan2010}. The dressing field approach provides a general unifying framework for  these works, and achieves the conceptual clarity philosophers of physics have been striving for \cite{Struyve2011, vanDam2011, Friederich2013}.

\section{From tetrad to metric formulation of General Relativity} 
\label{From tetrad to metric formulation of General Relativity} 
 
Einstein teaches us that gravitation is the dynamics of space-time, the base manifold itself. It deals with spatio-temporal degrees of freedom, not “inner” ones like in Yang-Mills-type gauge theories. In the most general case there exists a notion of \textit{torsion}, a concept absent in Yang-Mills theories. There are more possible invariants one can use in a Lagrangian due to index contractions impossible in Yang-Mills theories: the actual Lagrangian form for GR is not of Yang-Mills type. 

All this issues from the existence in gravitational theories of the \textit{soldering form}, also known as (co-)tetrad field, which realizes an isomorphism between the  tangent space at each point of space-time and  the Minkowski space \cite{Trautman}. The soldering form can be seen as the formal implementation of Einstein's ``happiest thought'', the \textit{Equivalence Principle}, which is the key specific physical feature distinguishing  the gravitational interaction from the three others (Yang-Mills) gauge interactions.

So, while  Yang-Mills fields are described by Ehresmann connections (principal connections) on a principal bundle, the gravitational field is described by both an Ehresmann connection, the Lorentz/spin connection, \textit{and} a soldering form. In $1977$, McDowell and Mansouri treated the concatenation of the  connection and of the soldering form  as a single gauge potential \cite{McDowell-Mansouri1977}. The mathematical foundation of this move is Cartan geometry \cite{Wise10, Wise09}: the third additional axiom defining a Cartan connection, and distinguishing it from a principal connection, defines an absolute parallelism on $\P$. This in turn induces, in simple cases, a soldering form \cite{Sharpe}. In other word, the geometry of the bundle $\P$ is much more tightly related to the geometry of the base manifold.  On can then convincingly argue  that Cartan geometry is a very natural framework for classical gravitational theories.

In the following we recast the tetrad formulation of GR in terms of the adequate Cartan geometry, and show that switching to the metric formulation can be seen as an application of the dressing field method. 
   
\subsection{Reduction of the Lorentz gauge symmetry}
\label{Reduction of the Lorentz gauge symmetry}
 
The relevant Cartan geometry is based on the Klein model $(G, H)$ given by $G=SO(1,3)\ltimes \RR^{1, 3}$, the Poincaré group, and $H=SO(1,3)$, the Lorentz group, so that the associated homogeneous space is $G/H=\RR^{1,3}$, the Minkowski space.  The infinitesimal Klein pair is $(\LieG, \LieH)$ with $\LieG=\so(1,3)\oplus \RR^{1,3}$ and $\LieH=\so(1,3)$. The principal bundle of this Cartan geometry is $\P\left(\M, SO(1,3)\right)$. The local Cartan connection and its curvature are the $1$-forms $ \varpi \in \bigwedge^1(\U, \LieG)$ and $\Omega \in \bigwedge^2(\U, \LieG)$, which can be written in matrix form
\begin{align*}
 \varpi&=
 \setlength\arraycolsep{5pt}
 \begin{pmatrix} A & \theta \\ 0 & 0  \end{pmatrix},
&
 \Omega&
 \setlength\arraycolsep{5pt}
 =\begin{pmatrix} R & \Theta \\ 0 & 0  \end{pmatrix}
 =\begin{pmatrix} dA +A\wedge A &  d\theta +A\wedge\theta \\ 0 & 0  \end{pmatrix},
\end{align*}
where $A \in \bigwedge^1(\U, \so)$ is the spin connection with Riemann curvature $2$-form $R$ and torsion $\Theta=D\theta$,  and $\theta \in \bigwedge^1(\U, \RR^{1, 3})$ is the soldering form.  In other words, this Cartan geometry is just the usual Lorentz geometry (with torsion). We can thus consider the  Cartan connection $\varpi$ as the gravitational gauge potential. The local gauge group is $\SO := \SO(1,3)$ and its action by an element $\gamma: \U\rarrow SO$, assuming the matrix form $\gamma=\begin{psmallmatrix} S & 0 \\ 0 & 1  \end{psmallmatrix}$, is 
\begin{align*}
\varpi^\gamma 
&=\gamma\-\varpi\gamma + \gamma\-d\gamma
\setlength\arraycolsep{8pt}
= \begin{pmatrix} S\-A S +S\-dS & S\-\theta \\ 0 & 0  \end{pmatrix},
\\
\Omega^\gamma 
&=\gamma\-\Omega\gamma
\setlength\arraycolsep{6pt}
=\begin{pmatrix}S\-RS &  S\-\Theta \\ 0 & 0   \end{pmatrix}.
\end{align*}
Given these geometrical data, the associated Lagrangian form of GR is given by,
\begin{align}
\label{Pal-Lagrangian}
L_\text{Pal}(A, \theta) =  \frac{-1}{32\pi \sG} \Tr\big(R\wedge * (\theta\wedge \theta^t)\big) =  \frac{-1}{32\pi \sG} \Tr\big(R\wedge * (\theta\wedge \theta^T\eta)\big),
 \end{align} 
with $\eta$ the metric of $\RR^{1, 3}$ and $\sG$ the gravitational constant. Given $\S=\int L$, variation w.r.t. $\theta$ gives Einstein's equation in vacuum and variation w.r.t. $A$ gives an equation for the torsion which in the vacuum vanishes (even in the presence of matter, the torsion does not propagate). 

\smallskip
Looking for a dressing field liable to neutralize the $\SO$-gauge symmetry, given the gauge transformation of the Cartan connection, the tetrad field $e={e^a}$ in the soldering form $\theta^a={e^a}_\mu dx^\mu$ is a natural candidate: $\theta^S=S\-\theta$ implies  $e^S=S\- e$, so that we define
\begin{align*}
\setlength\arraycolsep{4pt}
u=\begin{pmatrix} e &  0 \\ 0 & 1 \end{pmatrix} \quad \text{and we get }u^\gamma=\gamma\- u.
\end{align*}
Then  $u$ is a $\SO$-dressing field, and notice that its target group $G=GL$ is bigger than the structure group which happen to be also its equivariance group, $H=K=SO$.\footnote{While in the previous example we had $G=K=SU(2)\subset H=U(1)\times SU(2)$.} We can apply Prop.~\ref{P1} and construct the $\SO$-invariant composite fields,
\begin{align*}
\h\varpi 
&=u\-\varpi u+u\-du
\setlength\arraycolsep{6pt}
=  \begin{pmatrix} e\-Ae +e\-de & e\-\theta \\ 0 & 0 \end{pmatrix}
=: \begin{pmatrix} \Gamma & dx \\ 0 & 0  \end{pmatrix}, 
\\
\h\Omega &=u\-\Omega u
\setlength\arraycolsep{6pt}
=\begin{pmatrix} e\-Re & e\-\Theta\\ 0 & 0  \end{pmatrix}
=:\begin{pmatrix} \sR & T \\ 0 & 0  \end{pmatrix}, 
\end{align*}
where $\Gamma={\Gamma^\mu}_\nu={\Gamma^\mu}_{\nu, \rho}dx^\rho$ is the linear connection $1$-form on $\U\subset \M$, and
$\sR$ and $T$ are the Riemann curvature and torsion $2$-forms  written in the coordinates system $\{ x^\mu \}$ on $\U$. We have their explicit expressions as functions of the components of the dressed Cartan connection $\h\varpi$ on account of,
\begin{align*}
\h\Omega
=d\h\varpi +\h\varpi\wedge \h\varpi
\setlength\arraycolsep{4pt}
=\begin{pmatrix} d\Gamma & d^2x \\ 0 & 0\end{pmatrix} 
+ \begin{pmatrix} \Gamma\wedge \Gamma & \Gamma\wedge dx \\ 0 & 0\end{pmatrix} 
=\begin{pmatrix} d\Gamma +\Gamma\wedge\Gamma & \Gamma\wedge dx\\ 0 & 0\end{pmatrix}.
\end{align*}
We see clearly that if $\Gamma$ is symmetric on its lower indices, the torsion vanishes. 

A Cartan connection always induces a metric on the base manifold $\U\subset\M$ by $g(X, Y)=\eta\big(\theta(X), \theta(Y)\big)$, with $X, Y \in T_x\U$. In component this reads $g_{\mu\nu}={e_\mu}^a\eta_{ab}{e^b}_\nu$, or in index free notation $g=e^T\eta e$. Notice that by definition $g$ is $\SO$-gauge-invariant. It is easy to show that in this formalism, the metricity condition is necessarily satisfied: $\h Dg:=\nabla g=dg - \Gamma^Tg -g\Gamma =  -e^T\big( A^T\eta +\eta A \big)e = 0$. Therefore if $T=0$, $\Gamma$ is the Levi-Civita connection. 

Now by application of Prop.~\ref{Prop-Lagrangian} we see that the classic calculation that allows to switch from the $\SO$-gauge formulation to  the metric  formulation can be seen as an example of the dressing field method,
\begin{align*}
L_\text{Pal}(A, \theta)
&=\frac{-1}{32\pi \sG}\Tr\left( R\wedge *(\theta\wedge \theta^t) \right) 
=\frac{-1}{32\pi \sG}\Tr\big( \sR g\big)\wedge *(dx\wedge dx)
\\
&=\frac{1}{16\pi \sG} \sqrt{|g|} d^mx\   \sR{\textrm{{\scriptsize icc}}}
=:L_\text{EH}(\Gamma, g).
\end{align*}
The last  equation defines the  Einstein-Hilbert Lagrangian form, depending on the $\SO$-invariant composite fields $\Gamma$ and $g$.

\subsection{Residual symmetry}
\label{Residual symmetry}

The $\SO$-invariant fields $g$, $\h\varpi=(\Gamma, dx)$ and $\h\Omega=(\sR, T)$ belong to the \textit{natural geometry} of the base manifold $\M$, \textsl{i.e.} the geometry defined only in terms of its frame bundle and its associated vector bundles. The only residual transformations these fields can display are coordinates transformations. On the overlap of  two patches of coordinates $\{x^\mu\}$ and  $\{y^\mu\}$ in a trivializing open set $\U\subset \M$, the initial gauge fields $\varpi$ and $\Omega$, as differential forms, are well defined and invariant. But obviously $\theta=e dx=e' dy$ implies that the tetrad undergoes the transformation $e'=eG$, with $G={G^\mu}_\nu=\tfrac{\partial x^\mu}{\partial  y^\nu}$. The dressing fields then transforms as  $u'=uG$, with $G=\begin{psmallmatrix} G & 0 \\ 0 & 1 \end{psmallmatrix}$, so the composite fields have coordinates transformations, 
\begin{align*}
\h\varpi'
&={u'}\-\varpi u' +{u'}\-du'
=G\-\h\varpi G +G\-dG 
\\
&
\setlength\arraycolsep{6pt}
= \begin{pmatrix} G\-\Gamma G +G\-dG & G\-dx \\ 0 & 0 \end{pmatrix} 
=:\begin{pmatrix}\Gamma' & dy \\ 0 & 0 \end{pmatrix},
\\
\h \Omega'
&={u'}\-\Omega u'
=G\-\h \Omega G
\setlength\arraycolsep{6pt}
= \begin{pmatrix}G\-\sR G & G\- T \\ 0 & 0\end{pmatrix} 
=: \begin{pmatrix} \sR' & T' \\ 0 & 0 \end{pmatrix}, 
\\
g'&=e'^T\eta e'=G^TgG.
\end{align*}
This gives the well known transformations of the linear connection, of the metric, Riemann and torsion tensors under general changes of coordinates. Of course the  Lagrangian form, $L_\text{EH}$, is invariant.

\subsection{Discussion}  
\label{Discussion2}

The tetrad as a dressing field does not belong to the gauge group $\SO$ of the theory. So, strictly speaking, the invariant composite field $\h\varpi$ is not a gauge transformation of the Cartan connection $\varpi$. In particular this means that, contrary to what is sometimes said, $\Gamma$ is not a gauge transform of the Lorentz connection $A$. Indeed $\Gamma$ is an $\SO$-invariant $\gl$-valued $1$-form on $\M$, clearly it does not belong to the initial space of connections of the theory. Even if one considers that the gauge symmetry of GR are the coordinates changes, thinking of it as a gauge theory on the frame bundle $L\M$ with gauge group $\GL$, the tetrad ${e^a}_\mu$ still doesn't belong to $\GL$. So one cannot view $\Gamma$ and $A$ as gauge related. To obtain $A$ from $\Gamma$ one needs the bundle reduction theorem, which allows to reduce $L\M$ to the subbundle $\P\left(\M, SO(1,3)\right)$. To recover  $\Gamma$ from $A$, one needs to think in terms of the dressing field method.

\section{Conformal Cartan geometry, tractors and twistors} 
\label{Conformal Cartan geometry, tractors and twistors} 

In this Section we show how tractors and twistors, which are conformal calculi for torsionless manifolds \cite{Bailey-et-al94, Penrose-Rindler-vol2}, can be derived from the conformal Cartan geometry via the dressing field method. We thus start by a brief description of this geometry and then we deal with tractors and twistors.

\subsection{Conformal Cartan geometry in a nutshell} 
\label{Conformal Cartan geometry in a nutshell} 
  
A conformal Cartan geometry $(\P, \varpi)$ can be defined over $n$-manifolds $\M$ for any  $n\geq 3$ and signature $(r, s)$ thanks to the group $SO(r+1, s+1)$. We will admit that the base manifold is such that a corresponding spinorial version $(\b \P, \b\varpi)$ exists, based on the group $\Spin(r+1, s+1)$, so that we have the two-fold covering $\b\P \xrightarrow{2:1}  \P$. Since we seek to reproduce twistors in signature $(1, 3)$, as spinors corresponding to tractors, we are here interested in conformal Cartan geometry over $4$-manifolds, and thus take advantage of the accidental isomorphism $\Spin(2, 4)\simeq SU(2, 2)$.

We then treat in parallel the conformal Cartan geometry $(\P(\M, H), \varpi)$ modeled on the Klein model $(G, H)$ and its naturally associated vector bundle $E$, as well as the spinorial version $(\b\P(\M, \b H), \b\varpi)$ modeled on the Klein model $(\b G, \b H)$ and its naturally associated vector bundle $\sE$. For simplicity we designate them as the real and complex cases respectively. By dressing, the real case will yield tractors and the complex case will yield twistors.

In the real case, we have 
\begin{equation*}
G=PSO(2, 4)=\left\{ M \in GL_{6}(\RR)\  | \ M^T \Sigma M= \Sigma, \det{M}=1 \right\}/ \pm \id
\end{equation*}
with $\Sigma=\begin{psmallmatrix}  0 & 0 & -1 \\ 0 & \eta & 0 \\ -1 & 0 & 0 \end{psmallmatrix}$ the group metric, $\eta$ the flat metric of signature $(1, 3)$, and $H$ is a parabolic subgroup  comprising Lorentz, Weyl and conformal boost symmetries: it has the following matrix presentation \cite{Cap-Slovak09, Sharpe}, with $W := \RR^*_+$ (Weyl dilation group),
\begin{align*}
 H = K_0\, K_1=\left\{ 
 \setlength\arraycolsep{5pt}
 \begin{pmatrix} z &  0 & 0  \\  0  & S & 0 \\ 0 & 0 & z^{-1}  \end{pmatrix}\!  
 \begin{pmatrix} 1 & r & \tfrac{1}{2}rr^t \\ 0 & \1_4 & r^t \\  0 & 0 & 1\end{pmatrix}  
 \middle| \quad
 \begin{aligned}
&  z\in W,\\[-2pt]
&  S\in SO(1, 3), \\[-2pt]
& r\in \RR^{4*} 
 \end{aligned}
\right\},
\end{align*} 
where $K_0$  (resp. $K_1$) corresponds to the matrices on the left (resp. right) in the product. Clearly $K_0\simeq CO(1, 3)$ via $(S, z) \mapsto zS$, and $K_1$ is the abelian group of conformal boosts.  Here ${}^T\,$ is the usual matrix transposition, $r^t = (r \eta^{-1})^T$ stands for the $\eta$-transposition, and $\RR^{4*}$ is the dual of $\RR^4$.

The corresponding Lie algebras  $(\LieG, \LieH)$ are graded:  $[\LieG_i, \LieG_j] \subseteq \LieG_{i+j}$, $i,j=0,\pm 1$, with the abelian Lie subalgebras $[\LieG_{-1}, \LieG_{-1}] = 0 = [\LieG_1, \LieG_1]$. They decompose respectively as, $\LieG=\LieG_{-1}\oplus\LieG_0\oplus\LieG_1 \simeq \RR^4\oplus\co(1, 3)\oplus\RR^{4*}$, with $\co(1, 3) = \so(1, 3) \oplus\RR$,  and $\LieH=\LieG_0\oplus\LieG_1 \simeq \co(1, 3)\oplus\RR^{4*}$. In matrix notation we have,
\begin{align*}
\mathfrak{g} = \left\{ 
\setlength\arraycolsep{5pt}
\begin{pmatrix} \epsilon &  \iota & 0  \\  \tau  & s & \iota^t \\ 0 & \tau^t & -\epsilon  \end{pmatrix} 
\middle| \quad
 \begin{aligned}
&  (s-\epsilon\1_4)\in \mathfrak{co}(1, 3),\\[-2pt]
&  \tau\in\mathbb{R}^4,\\[-2pt]
& \iota\in\mathbb{R}^{4*}
 \end{aligned}
\right\} 
\supset
\LieH = \left\{
\setlength\arraycolsep{5pt}
 \begin{pmatrix} \epsilon &  \iota  & 0  \\  0  & s & \iota^t \\ 0 & 0 & -\epsilon  \end{pmatrix} 
 \right\}.
\end{align*} 
The graded structure of the Lie algebras is automatically handled by the matrix commutator.

\smallskip
In order to introduce the complex case, let us first consider the canonical isomorphism of vector spaces between Minkowski space $\mathbb{R}^4$ and hermitian  $2\times2$  matrices $\Herm(2, \CC)=\left\{ M \in GL_2(\CC)\ | \ M^*=M  \right\}$, where ${}^*$ means trans-conjugation: $\mathbb{R}^4 \rarrow \Herm(2, \CC)$, $x\mapsto \b x = x^a \sigma_a$ ($\sigma_0 = \1_2$ and $\s_{i=\{1, 2,3\}}$ are the Pauli matrices). There is a corresponding double covering group morphism $SL(2, \CC) \xrightarrow{2:1} SO(1, 3)$, $\b S\mapsto S$ (so that $S\- x \mapsto \b S\- \b x \b S^{-1*}$ and $x^t S \mapsto \b S^*\b x^t \b S$), and its associated Lie algebra isomorphism $\so(1, 3) \simeq \sl(2, \CC)$ is denoted by $s \mapsto \b s$. In the following, the bar notation will relate the “real” and “complex” cases in a natural way by using same letters, so generalizing the above maps.

For the complex case, we have then  $\b G = SU(2,2) \simeq \Spin(2,4)$, which is the group preserving the metric $\b\Sigma=\begin{psmallmatrix} 0 & \1_2 \\ \1_2 & 0 \end{psmallmatrix}$,  and $\b H$ is given in matrix notation by
\begin{align}
\label{IsoH-bH}
\b H=\b K_0 \b K_1 &:= \left\{ 
\setlength\arraycolsep{3pt}
\begin{pmatrix}[1.2] z^{\sfrac{1}{2}} {\b S}^{-1*} & 0 \\ 0 & z^{-\sfrac{1}{2}} \b S\  \end{pmatrix}
\begin{pmatrix}[1.2] \1_2 & -i \b r \\ 0 & \1_2 \end{pmatrix} 
\middle| \quad
 \begin{aligned}
&  z\in W,\ \b S\in SL(2, \CC),\\[-2pt]
&  \b r\in \Herm(2, \CC)
 \end{aligned}
\right\}.
\end{align}
There is a double covering $\b H \xrightarrow{2:1} H$ which reduces to a double covering  $\b K_0 \xrightarrow{2:1} K_0$ and a natural isomorphism $\b K_1 \simeq K_1$. Using the bar notation, the Lie algebra isomorphism $\so(2, 4)=\LieG \rarrow \su(2, 2)=\b \LieG$ is explicitly given by 
\begin{align}
\label{LieAlg-Iso}
\b\LieG &=\b\LieG_{-1}+\b\LieG_0+\b\LieG_1
= \left\{ 
\setlength\arraycolsep{4pt}
\begin{pmatrix} -(\b s^* -\tfrac{\epsilon}{2}\1_2) &  -i\b\iota \\  i\b\tau  & \b s-\tfrac{\epsilon}{2}\1_2  \end{pmatrix}
\middle| \quad
 \begin{aligned}
&  \epsilon\in\RR, \b s\in \sl(2,\CC)\\[-2pt]
&  \b \tau, \b\iota \in \Herm(2, \CC)
 \end{aligned}
 \right\}
\notag\\
&\supset
\b\LieH = \b\LieG_0+\b\LieG_1.
\end{align}

Once given two  Cartan bundles such that $\b\P(\M, \b H) \xrightarrow{2:1} \P(\M, H)$, we endow $\P(\M, H)$ with a conformal Cartan connection whose local representative on $\U \subset \M$ is $\varpi  \in \bigwedge^1(\U , \LieG)$, with curvature $\Omega\in \bigwedge^2(\U, \LieG)$. In matrix presentation, one has
\begin{align*}
\setlength\arraycolsep{5pt}
\varpi =\begin{pmatrix} a & P & 0 \\ \theta & A & P^t \\0 & \theta^t & -a \end{pmatrix} 
\quad \text{and} \quad  
\Omega=d\varpi+\varpi^2=\begin{pmatrix} f & C & 0 \\ \Theta & W & C^t \\0 & \Theta^t & -f \end{pmatrix}.
\end{align*}
In the same way, $\b\P(\M, \b H)$ is endowed with a spinorial Cartan connexion
\begin{align*}
\setlength\arraycolsep{5pt}
\b\varpi=\begin{pmatrix}[1.2]  -( \b A^* - \tfrac{a}{2}\1_2 )  &  -i\b P \\ i\b \theta & \b A -\tfrac{a}{2}\1_2  \end{pmatrix} 
\quad \text{ and }  \quad 
\b\Omega=\begin{pmatrix}[1.2]  -( \b W^* - \tfrac{f}{2}\1_2 )  &  -i\b C \\ i\b \Theta & \b W -\tfrac{f}{2}\1_2  \end{pmatrix}.
\end{align*}

The soldering part of $\varpi$ is $\theta=e\cdot dx$, \textit{i.e.}  $\theta^a:={e^a}_\mu dx^\mu$.\footnote{Notice that from now on we shall make use of  ``$\cdot$'' to denote Greek indices contractions, while Latin indices contraction is naturally understood from matrix multiplication.} Denote by $g$ the metric of signature $(1, 3)$ on $\M$ induced from $\eta$ via  $\varpi$ according to $g(X, Y):=\eta\left( \theta(X), \theta(Y)\right)=\theta(X)^T\eta \theta(Y)$, or in a way more familiar to physicists $g:=e^T\eta e$, so that $g_{\mu\nu}={e_\mu}^a\eta_{ab} {e^b}_\nu$. The action of $\H$ on $\varpi$ induces, through $\theta$, a conformal class of metrics $c:=[g]$ on $\M$. But $(\P, \varpi)$ is not equivalent to $(\M, c)$. Nevertheless, there is a distinguished choice, the so-called \emph{normal} conformal Cartan connection $\varpi_\n$, which is unique in satisfying the conditions $\Theta=0$ and ${W^a}_{bad}=0$ (which in turn, through the Bianchi identity, implies $f=0$), so that $(\P, \varpi_\n)$ is indeed equivalent to a conformal manifold $(\M, c)$. 

Still, it would be hasty to  identify $A$ in $\varpi$ or $\varpi_\text{\tiny{N}}$ with the Lorentz connection one is familiar with in physics, and by a way of consequence to take $R:=dA+A^2$ and $P$ as the Riemann and Schouten tensors. Indeed,  contrary to expectations, $A$ is invariant under Weyl rescaling and neither $R$ nor $P$ have the well-known Weyl transformations. It turns out that one recovers the spin connection and the mentioned associated tensors only after a dressing operation, as shown in \cite{Attard-Francois2016_I}. 

Using the natural representation of $H$ on $\RR^6$, we can introduce the associated vector bundle $E=\P \times_H \RR^6$. A section of $E$ is a $H$-equivariant map on $\P$ whose local expression is $\vphi: \U \subset \M \rarrow \RR^6$, given explicitly as column vectors
\begin{align*}
\vphi=\begin{pmatrix} \rho\\[1mm] \ell \\ \s  \end{pmatrix}, \quad \text{ with }  \ell=\ell^a \in \RR^4, \text{ and } \rho, \s \in \RR.
\end{align*}
The covariant derivative induced by the Cartan connection is $D\vphi=d\vphi+\varpi\vphi$,  with $D^2 \vphi= \Omega \vphi$. The group metric $\Sigma$ defines an invariant bilinear form on sections of $E$: for any $\vphi, \vphi' \in \Gamma(E)$, one has $\langle \vphi, \vphi'  \rangle= \vphi^T \Sigma \vphi'= -\s\rho' + \ell^T \eta \ell' - \rho\s'$. The covariant derivative $D$ preserves this bilinear form since $\varpi$ is $\LieG$-valued: $D\Sigma=d\Sigma + \varpi^T \Sigma + \Sigma \varpi=0$.

We now follow the same line of constructions in the complex case, using the natural representation $\CC^4$ of $\b H$ to define the associated vector bundle $\sE=\b\P \times_{\b H}\CC^4$. A section of $\sE$ is a $\b H$-equivariant map on $\b\P$ whose local expression is $\psi: \U \subset \M \rarrow \CC^4$ given as
\begin{align*}
\psi=\begin{pmatrix} \pi\\ \omega  \end{pmatrix}, \quad \text{ with }  \pi, \omega \in \CC^2 \text {dual Weyl spinors}.
\end{align*}
The covariant derivative is now $\b D\psi=d\psi+\b\varpi\psi$, with $\b D^2 \psi=\b \Omega \psi$. The group metric $\b\Sigma$ defines an invariant bilinear form on sections of $\sE$: for any $\psi, \psi' \in \Gamma(\sE)$, one has $\langle \psi, \psi'  \rangle= \psi^* \b\Sigma \psi' =\pi^*\omega'+ \omega^*\pi'$. Again, the covariant derivative $\b D$ preserves  this bilinear form.

The gauge groups $\H=\K_0\K_1$ and $\b\H = \b\K_0\b\K_1$ act on the gauge variables, with $\gamma \in \H$ and $\b\gamma \in \b\H$, as
\begin{align*}
\varpi^\gamma &= \gamma\- \varpi \gamma + \gamma\- d\gamma,  
&
\vphi^\gamma &=\gamma\- \vphi, 
&
\b\varpi^{\b\gamma} &= {\b\gamma}\- \b\varpi {\b\gamma} + {\b\gamma}\- d{\b\gamma},  
&
\psi^{\b\gamma} &={\b\gamma}\-\psi. 
\end{align*}
This induces the actions $\Omega^\gamma = \gamma\- \Omega \gamma$, $\b\Omega^{\b\gamma} = {\b\gamma}\- \b\Omega {\b\gamma}$, $(D\vphi)^\gamma = \gamma\- D\vphi$, and $(\b D\psi) ^{\b\gamma} ={\b\gamma}\-\b D\psi$. Given $\gamma_0\in \K_0$, the soldering part of the gauge transformed Cartan connection $\varpi^{\gamma_0}$ is $\theta^{\gamma_0}=zS\- \theta$, so that the metric induced by $\varpi^{\gamma_0}$ is $g'=z^2g$. On the other hand, $\theta^{\gamma_1}=\theta$ for $\gamma_1 \in \K_1$. So, as mentioned above, the action of the gauge group induces a conformal class of metric $c$ on $\M$.

\subsection{Tractors and twistors: constructive procedure via dressing} 
\label{Tractors and twistors: constructive procedure via dressing} 

It has been noticed that tractor and twistor vector bundles are associated to the conformal Cartan bundle, and that tractor and twistor connections are related to the conformal Cartan connection \cite{Bailey-et-al94, Friedrich77}. However as it stands, the gauge transformations above show that $\vphi$ is not a tractor and that $\psi$ is not a twistor. It turns out that to recover tractors and twistors one needs to erase the conformal boost symmetry $\K_1 \simeq \b\K_1$. We outline the procedure below and give the important results. Details can be found in \cite{Attard-Francois2016_I, Attard-Francois2016_II}. 

Given the decompositions $H=K_0 K_1$ and $\b H=\b K_0 \b K_1$, the most natural choice of dressing field to erase the conformal boost gauge symmetry is $u_1 :\U\rarrow K_1$ in the real case and its corresponding element $\b u_1 : \U \rarrow \b K_1 \simeq K_1$ in the complex case, given by
\begin{align*}
 u_1 &
 \setlength\arraycolsep{5pt}
 =\begin{pmatrix}  1 & q & \tfrac{1}{2}qq^t \\ 0 & \1_4 & q^t \\ 0 & 0 & 1 \end{pmatrix},
 & 
 \b u_1 &
  \setlength\arraycolsep{5pt}
=\begin{pmatrix} \1_2 & -i \b q \\ 0 & \1_2 \end{pmatrix}.
\end{align*}
It turns out that $u_1$ can be defined via the ``gauge-like'' constraint  $\Sigma(\varpi^{u_1}):=\Tr(A^{u_1} - a^{u_1})=-na^{u_1}=0$. Indeed, this gives the equation $a-q\theta=0$, which once solved for $q$ gives $q_a=a_\mu {e^\mu}_a$, or in index free notation $q=a\cdot e\-$.\footnote{Beware of the fact that in this index free notation $a$ is the set of components of the $1$-form $a$. This should be clear from the context.} Now, from $\varpi^{\gamma_1}$ one finds that $q^{\gamma_1}=a^{\gamma_1}\cdot (e^{\gamma_1})\-=(a-re)\cdot e\-=q-r$. One then checks easily that the constraint $\Sigma(\varpi^{u_1})=0$ is $\K_1$-invariant and that $u_1$ is a dressing field for $K_1$: from $q^{\gamma_1}=q-r$ one shows that $ u_1^{\gamma_1}=\gamma_1\- u_1$. In the same way, one has $\b u_1^{\b\gamma_1}=\b\gamma_1\- \b u_1$.

With these $\K_1$-dressing fields, we can apply (the local version of) Prop.~\ref{P1} and form the $\K_1 \simeq \b\K_1$-invariant composite fields in the real and complex cases:
\begin{align*}
\varpi_1 &:=\varpi^{u_1}
=u_1 \- \varpi u_1 + u_1\-du_1
\setlength\arraycolsep{4pt}
=\begin{pmatrix} 0 & P_1 & 0 \\ \theta & A_1 & P_1^t \\0 & \theta^t & 0 \end{pmatrix}, 
&
\b\varpi_1 &=\b\varpi^{\b u_1} 
\setlength\arraycolsep{4pt}
=\begin{pmatrix}[1.2]  - \b A_1^*   &  -i\b P_1 \\ i\b \theta & \b A_1  \end{pmatrix}
\notag\\
\Omega_1&:=\Omega^{u_1}=u_1\-\Omega u_1=d\varpi_1+\varpi_1^2,
&
\b\Omega_1&=\b\Omega^{\b u_1}=\b u_1\-\b\Omega \b u_1,
\end{align*}
\begin{align*}
\vphi_1 &:=u_1\-\vphi, 
\\
D_1\vphi_1 &=d\vphi_1+\varpi_1 \vphi_1
=\begin{pmatrix} d\rho_1 +P_1 \ell_1\\[1mm] d\ell_1+A_1\ell_1 + \theta \rho_1 + P_1^t \s \\ d\s+\theta^t \ell_1  \end{pmatrix}
=\begin{pmatrix} \nabla\rho_1 +P_1 \ell_1\\ \nabla\ell_1 + \theta \rho_1 + P_1^t \s \\ \nabla\s+\theta^t \ell_1  \end{pmatrix}, 
\notag
\end{align*}
\begin{align*}
\psi_1 &:=\b u_1\-\psi,
\\
\b D_1\psi_1 &=d\psi_1+\b\varpi_1 \psi_1
=\begin{pmatrix}[1.2] d\pi_1  -\b A_1^*\pi_1 -i\b P_1 \omega_1 \\ d\omega_1 + \b A_1\omega_1 + i\b\theta \pi_1 \end{pmatrix}
=\begin{pmatrix}[1.2] \b\nabla\pi_1 -i\b P_1 \omega_1 \\ \b\nabla\omega_1 + i\b\theta \pi_1  \end{pmatrix}, 
\notag 
\end{align*}
with obvious notations. As expected, $D_1^2 \vphi_1 = \Omega_1 \vphi_1$ and ${\b D_1}^2 \psi_1 = \b\Omega_1 \psi_1$. Notice also that $f_1=P_1\w\theta$ is the antisymmetric part of the tensor $P_1$.  

We claim that $\vphi_1$ is a tractor and that the covariant derivative $D_1$ is a ``generalized'' tractor connection \cite{Bailey-et-al94}. In the same way, we assert that $\psi_1$ is a twistor and that the covariant derivative $\b D_1$ is a generalized twistor connection \cite{Penrose-Rindler-vol2}. Both assertions are supported by the analysis of the residual gauge symmetries.

\paragraph{Residual gauge symmetries.} Being by construction $\K_1 \simeq \b\K_1$-invariant, the composite fields collectively denoted by $\chi_1$ are expected to display $\K_0$-residual and $\b\K_0$-residual gauge symmetries. The group $K_0$  breaks down as a direct product of the Lorentz and Weyl groups, $K_0 = SO(1, 3) W$, and in the same way, $\b K_0= SL(2, \CC) W$, with respective matrix presentations
\begin{align}
\label{K0 matrix}
K_0 &=
\left\{ 
\sS \sZ :=
\setlength\arraycolsep{4pt}
\begin{pmatrix} 1 &  0 & 0  \\  0  & S & 0 \\ 0 & 0 & 1  \end{pmatrix}\!
\begin{pmatrix} z &  0 & 0  \\  0  & \1_4 & 0 \\ 0 & 0 & z^{-1}  \end{pmatrix}
\bigg|
 \ z\in W,\ S\in SO(1, 3)
\right\}
\\
\label{K0 bar matrix}
\b K_0 &=
\left\{ 
\b\sS \b \sZ :=
\setlength\arraycolsep{4pt}
\begin{pmatrix}  {\b S}^{-1*} & 0 \\ 0 &  \b S\  \end{pmatrix}\!
\begin{pmatrix} z^{\sfrac{1}{2}} & 0 \\ 0 & z^{-\sfrac{1}{2}}  \end{pmatrix}
\bigg|
 \ z\in W,\ \b S\in SL(2, \CC)
\right\}
\end{align}

 We focus on Lorentz symmetry first, then only bring our attention to Weyl symmetry. In the following, we will use the above matrix presentations $\sS$ and $\b\sS$ for elements of the Lorentz gauge group $\SO$ and the $SL(2,\CC)$-gauge group $\SL$. The residual gauge transformations of the composite fields under $\SO$  is inherited from that of the dressing field $u_1$. Using $\varpi^{\gamma_0}$ to compute $q^{ \sS}=a^{ \sS}\cdot (e^{ \sS})\-=qS$, one easily finds that $u_1^\sS=\sS\- u_1 \sS$, and correspondingly, $\b u_1^{\b{\sS}}={\b{\sS}}\- \b u_1 \b{\sS}$. This is a local instance of Prop.~\ref{Prop2}, which then allows to conclude that the composite fields $\chi_1$ are \emph{genuine} gauge fields (see Section~\ref {The composite fields as genuine gauge fields}), w.r.t. Lorentz gauge symmetry. Hence, from Cor.~\ref{Cor3} follows that the residual $\SO$-gauge  and $\SL$-gauge transformations are:
\begin{align}
\label{CompFields_1_S}
\varpi_1^\sS&={\sS}\-  \varpi_1 \sS  + { \sS}\- d \sS
\setlength\arraycolsep{6pt}
= \begin{pmatrix} 0 & P_1 S &  0 \\ S\-\theta  & S\-A_1S +S\-dS & S\-P^t   \\  0  & \theta^t S &  0 \end{pmatrix},
\end{align}
\begin{align}
\label{CompFields_1_S_bar}
\b\varpi_1^{\b{\sS}}={\b{\sS}}\-  \b\varpi_1 {\b{\sS}}  + {\b{\sS}}\- d {\b{\sS}}
\setlength\arraycolsep{5pt}
=\begin{pmatrix}[1.2] -\left(  \b S^*\b A_1 {\b S}^{-1*} + d\b S^*{\b S}^{-1*} \right) &  -i\ \b S^* \b P_1 \b S \\  i\ {\b S}\-\b\theta {\b S}^{-1*} &  {\b S}\-\b A_1 \b S + {\b S}\- d\b S &  \end{pmatrix}, \notag\\[2mm]
\end{align}
and
\begin{align}
\label{CompFields_2_S}
 \Omega_1^\sS &={ \sS}\- \Omega_1  \sS,  
 &
\vphi_1^\sS &= {\sS}\- \vphi_1,
 &
  (D_1\vphi_1)^\sS &= {\sS}\- D_1\vphi_1, 
 \\
 \label{CompFields_2_S_bar}
\b\Omega_1^{\b{\sS}} &={\b{\sS}}\- \b\Omega_1 {\b{\sS}},   
&
\psi_1^{\b{\sS}} &= {\b{\sS}}\- \psi_1,
&
(\b D_1\psi_1)^{\b{\sS}} &={\b{\sS}}\- \b D_1\psi_1.
\end{align}
 See \cite{Attard-Francois2016_I, Attard-Francois2016_II} for details. Notice  that $\vphi_1$ and $\psi_1$ transform as sections of the $SO(1,3)$-associated bundle $E_1=E^{u_1}=\P\times_{SO} \RR^{6}$ and the $SL(2, \CC)$-associated bundle $\sE_1=\sE^{\b u_1} =\b\P\times_{SL} \CC^4$ respectively.

\smallskip
We repeat the exact same procedure to analyze the Weyl gauge symmetry, using again the matrix notations defined in \eqref{K0 matrix} and \eqref{K0 bar matrix} for $\sZ$ in the Weyl group $\W \subset \K_0$ and $\b\sZ$ in its complex counterpart $\b\W \subset \b\K_0$. We first compute the action of $\W$ on the dressing field: using $\varpi^{\gamma_0}$ to compute $q^\sZ=a^\sZ\cdot (e^\sZ)\-$, one easily finds that $u_1^\sZ=\sZ\- u_1 C(z)$, where $C: W \rarrow K_1 W \subset H$ is defined by
\begin{align}
\label{WeylGT_u_1}
C(z)&:=k_1(z) \sZ
\setlength\arraycolsep{5pt}
=\begin{pmatrix} 1 & \Upsilon & \tfrac{1}{2} \Upsilon ^2 \\ 0 & \1_4 & \Upsilon^t \\ 0 & 0 & 1 \end{pmatrix} 
\begin{pmatrix} z & 0 & 0 \\ 0 & \1_4 & 0 \\ 0 & 0 & z\- \end{pmatrix}
= \begin{pmatrix}[1.2] z & \Upsilon &  \tfrac{z\-}{2} \Upsilon ^2 \\ 0 & \1_4 & z\-\Upsilon^t \\ 0 & 0 & z\-  \end{pmatrix} 
\end{align}
where explicitly $\Upsilon=\Upsilon_a=\Upsilon_\mu {e^\mu}_a$, with $\Upsilon_\mu:=z\-\d_\mu z$, and $\Upsilon^2=\Upsilon_a \eta^{ab} \Upsilon_b$. The corresponding complex case is ${\b u_1}^{\b \sZ}={\b \sZ}\- \b u_1 \b C(z)$, where $\b C: W \rarrow \b K_1 W   \subset \b H$ is defined by, with $\b \Upsilon = \Upsilon_a \sigma^a$,
\begin{align}
\label{WeylGT_u_1_2}
\b C(z) &:=\b k_1(z)\b \sZ 
\setlength\arraycolsep{4pt}
=\begin{pmatrix} \1_2 & -i\b \Upsilon \\ 0 & \1_2 \end{pmatrix} 
\begin{pmatrix}z^{\sfrac{1}{2}}\1_2 & 0 \\ 0 &  z^{-\sfrac{1}{2}}\1_2 \end{pmatrix}
= \begin{pmatrix}[1.2] z^{\sfrac{1}{2}} \1_2 &  -i\ z^{-\sfrac{1}{2}}\b \Upsilon \\ 0 & z^{-\sfrac{1}{2}}\1_2 \end{pmatrix}. 
\end{align}
The map $C$ is not a group morphism, $C(z)C(z')\neq C(zz')$, but is a local instance of a  $1$-$\alpha$-cocycle satisfying Prop.~\ref{alpha_cocycle}:  $C(zz')=C(z'z)=C(z')\ {\sZ'}\- C(z) \sZ'$. Under a further $\W$-gauge transformation and due to $e^\sZ=ze$, one has $k_1(z)^{\sZ'}= {\sZ'}\- k_1(z)\sZ'$, which implies $C(z)^{\sZ'}= {\sZ'}\- C(z) \sZ'$. So, if $u_1$ undergoes a further $\W$-gauge transformation $\sZ'$, we get $\big(u_1^\sZ\big)^{\sZ'}=\big(\sZ^{\sZ'}\big)\- u_1^{\sZ'} C(z)^{\sZ'}=\sZ\-\ {\sZ'}\- u_1C(z')\ {\sZ'}\- C(z) \sZ'= (\sZ\sZ')\- u_1 C(zz')$. \emph{Mutadis mutandis}, all this is true for $\b C$ in \eqref{WeylGT_u_1_2} and for $\b u_1$ as well. We have then a  well-behaved action of the gauge groups $\W$ and $\b\W$ in the real and complex cases.
  
From this we conclude that the composite fields $\chi_1$ are  instances of generalized gauge fields described in Section~\ref{The composite fields as twisted-gauge fields}. By Prop.~\ref{P5},  the residual $\W$-gauge and $\b\W$-gauge transformations  are $\varpi_1^\sZ=C(z)\- \varpi_1 C(z) + C(z)\-dC(z)$ and $\b\varpi_1^{\b \sZ}={\b C(z)}\- \b\varpi_1 \b C(z) + {\b C(z)}\-d\b C(z)$, explicitly given by
\begin{align}
\varpi_1^\sZ&=
\setlength\arraycolsep{8pt}
\begin{pmatrix}[1.2] 0 & z\-\big( P_1 + \nabla\Upsilon - \Upsilon\theta \Upsilon+ \tfrac{1}{2} \Upsilon^2 \theta^t \big)  &  0  \\  z\theta &  A_1 + \theta\Upsilon- \Upsilon^t\theta^t & * \\0 & z\theta^t & 0 \end{pmatrix},  
\label{varpi_1_Z}    \\            
\b\varpi_1^{\b \sZ}&=
\setlength\arraycolsep{6pt}
\begin{pmatrix}[1.2] -\b A^*_1 -(\b\Upsilon\b\theta)_0\ \  &  -i\ z\-\left[\b P_1  + \left(d\b\Upsilon -\b\Upsilon \b A_1 - \b A_1^* \b\Upsilon\right) -\b\Upsilon\b\theta\b\Upsilon \right]    \\ i\ z\b\theta  &   \b A_ 1 + (\b\theta\b\Upsilon)_0  \end{pmatrix},
\label{varpi_1_Z_bar} 
 \end{align}
 where $(\b\theta\b\Upsilon)_0$ is the $\sl(2, \CC)$ part of $\b\theta\b\Upsilon=(\b\theta\b\Upsilon)_0 + \tfrac{\Upsilon\theta}{2}\1_2$. 
And (see  \cite{Attard-Francois2016_I, Attard-Francois2016_II} for details): 
 \begin{align}         
\Omega_1^\sZ &= C(z)\- \Omega_1 C(z)    
&                     
\b\Omega_1^{\b \sZ}&={\b C(z)}\- \b\Omega_1 \b C(z)      
\label{Omega_1_Z}   
\end{align}
 \begin{align}   
 \vphi_1^\sZ &=C(z)\- \vphi_1 
 = \begin{pmatrix} z\-\left( \rho_1 - \Upsilon\ell_1 + \tfrac{\s}{2} \Upsilon^2 \right) \\[2mm] \ell_1 - \Upsilon^t \s \\ z\s \end{pmatrix}, 
 &
 (D_1\vphi_1)^\sZ &= C(z)\- D_1\vphi_1,  
 \label{Tractor_Connection_1}   
 \\      
\psi_1^{\b \sZ} &=\b C(z)\- \psi_1 
= \begin{pmatrix}[1.2] z^{-\sfrac{1}{2}}\left( \pi_1 + i\b\Upsilon \omega_1\right) \\ z^{\sfrac{1}{2}} \omega_1 \end{pmatrix}, 
&
(\b D_1\psi_1)^{\b \sZ} &= \b C(z)\- \b D_1\psi_1.  \label{Twistor_Connection_1}
\end{align}

From \eqref{varpi_1_Z}, we see that $A_1$ exhibits the known Weyl transformation for the Lorentz connection, and  $P_1$ transforms as the Schouten tensor (in an orthonormal basis). But, actually, the former genuinely reduces to the latter only when one restricts to the  dressing of the \emph{normal} Cartan connection $\varpi_{\n, 1}$, so that $A_1$ is a function of $\theta$ and $\sP_1=P_1(A_1)$ is the symmetric Schouten tensor. So $f_1$ vanishes and we have,
\begin{align}
\Omega_{\text{\tiny N},1} &=d\varpi_{\text{\tiny N},1}+\varpi_{\text{\tiny N},1}^2
\setlength\arraycolsep{8pt}
=\begin{pmatrix} 0 & \sC_1 & 0 \\ 0 & \sW_1 & \sC_1^t \\0 & 0 & 0 \end{pmatrix}, 
\label{tractor_curv}
\\
\Omega_{\text{\tiny N},1}^\sZ &=C(z)\- \Omega_{\text{\tiny N},1}C(z)
\setlength\arraycolsep{8pt}
=\begin{pmatrix} 0  &   z\-\left(\sC_1 - \Upsilon \sW_1\right)  & 0  \\ 0 & \sW_1 & * \\ 0 & 0 & * \end{pmatrix}, 
\\
\b\Omega_{\text{\tiny N},1} &=d\b\varpi_{\text{\tiny N},1}+\b\varpi_{\text{\tiny N},1}^2
\setlength\arraycolsep{6pt}
=\begin{pmatrix}[1.2]  -\b \sW_1^*   &  -i\b \sC_1 \\ 0 & \b \sW_1 \end{pmatrix}, 
\label{twistor_curv}
\\
\b\Omega_{\text{\tiny N},1}^{\b \sZ} &=\b C(z)\- \b\Omega_{\text{\tiny N},1}\b C(z)
\setlength\arraycolsep{6pt}
=\begin{pmatrix}[1.2]  -\b \sW_1^*   &  -i \ z\-\!\left(\b \sC_1- \b\Upsilon\b \sW_1 - \b \sW_1^*\b\Upsilon \right) \\ 0 & \b \sW_1 \end{pmatrix}.  
\end{align}
We  see that $\sC_1=\nabla \sP_1$ is the Cotton tensor, and indeed transforms as such, while $ \sW_1$ is the invariant Weyl tensor. 

From $\vphi_1^\sZ$ in \eqref{Tractor_Connection_1}, we see that  the dressed section $\vphi_1$ is a section of the $C(W)$-twisted vector bundle $E_1\!=E^{u_1}\!=\P \times_{C( W)}\RR^{n+2}$ (see \eqref{C twisted vector bundle}). But this same relation is also precisely the defining Weyl transformation of a tractor field as derived in \cite{Bailey-et-al94}.  Then $E_1$ is the so-called \emph{standard tractor bundle}. Since $C(z) \in K_1 W \subset H$, we have $(C(z)\-)^T \Sigma C(z)\-=\Sigma$. So the bilinear form on $E$ defined by the group metric $\Sigma$  is also defined on $E_1$: $\left\langle \vphi_1, \vphi'_1  \right\rangle= \vphi_1^T \Sigma \vphi'_1$. This is otherwise known as the \emph{tractor metric}. Furthermore, $(D_1\vphi_1)^\sZ$ in \eqref{Tractor_Connection_1} shows that the operator $D_1:=d + \varpi_1$  is a generalization of the \emph{tractor connection} \cite{Bailey-et-al94, Curry-Gover2015}. The term ``connection'', while not inaccurate, could hide the fact that $\varpi_1$ is no more a geometric connection w.r.t. Weyl symmetry. So we shall prefer to call $D_1$ a  generalized tractor \emph{covariant derivative}. The standard tractor covariant derivative  is recovered by restriction to the dressing of the normal Cartan connection, $D_{\text{\tiny N},1}=d + \varpi_{\text{\tiny N},1}$, and $\Omega_{\text{\tiny N},1}$ in \eqref{tractor_curv} is known as the \emph{tractor curvature}.  

In the same way, $\psi_1^{\b \sZ}$ in \eqref{Twistor_Connection_1} shows that the dressed section $\psi_1$ is a section of the $\b C(W)$-twisted vector bundle $\sE_1\!=\sE^{\b u_1}\!=\b\P \times_{\b C( W)}\CC^4$. This same relation is also, modulo the $z$ factors, the defining Weyl transformation of a local twistor as given by Penrose \cite{Penrose-Rindler-vol2}. So  $\sE_1$ is identified with the \emph{local twistor bundle}. It is endowed with a bilinear form defined by the group metric $\b \Sigma$ of $SU(2, 2)$:  $\left\langle \psi_1, \psi'_1  \right\rangle= \psi_1^T \b\Sigma \psi'_1$. It is well-defined since, in view of $\b C(z) \in \b K_1  W \subset \b H$, we have $(C(z)\-)^* \b\Sigma C(z)\- =\b\Sigma$. In the twistor literature, the quantity $\tfrac{1}{2}\left\langle \psi_1,  \psi_1 \right\rangle$ is known as the \emph{helicity} of the twistor field $\psi_1$ \cite{Penrose-McCallum72, Penrose99}. Also, $(\b D_1\psi_1)^{\b \sZ}$ in \eqref{Twistor_Connection_1} shows that the operator $\b D_1:=d + \b\varpi_1$ is a generalization of the \emph{twistor connection}. For the reason stated above, we shall prefer to call $\b D_1$ a  generalized twistor covariant derivative. The usual twistor covariant derivative is recovered by restriction to the  normal case, $\b D_{\text{\tiny N},1}=d + \b\varpi_{\text{\tiny N},1}$, and $\b\Omega_{\text{\tiny N},1}$ in \eqref{twistor_curv} is known as the \emph{twistor curvature}.

Remark that the actions of the Lorentz/$SL(2,\CC)$ and Weyl gauge groups on the composite fields $\chi_1$ commute. In the real case for instance, we have  $\sS^\W=\sS$ so that  $\big(\chi_1^\SO\big)^\W =\big(\chi_1^\sS\big)^\W =\big( \chi_1^\W \big)^{\sS^\W}=\big( \chi_1^{C(z)}\big)^\sS=\chi_1^{C(z)\sS}$. But we also have $C(z)^\SO={\sS}\- C(z)\sS$, so we get $\big( \chi_1^\W \big)^\SO =\big( \chi_1^{C(z)}\big)^\SO =\big( \chi_1^\SO\big)^{C(z)^\SO} =\big(\chi_1^\sS \big)^{{\sS}\- C(z)\sS}=\chi_1^{C(z)\sS}$. Our notations for the tractor and twistor bundles can then be refined to reflect this: $E_1=\P\times_{ C(W)\cdot SO} \RR^6$ and $\sE_1=\P\times_{ \b C(W)\cdot SL} \CC^4$.

\smallskip
Following the ending considerations of Section~\ref{The composite fields as genuine gauge fields}, the fact that the composite fields $\varpi_1, \vphi_1$ are genuine Lorentz-gauge fields satisfying \eqref{CompFields_1_S} and \eqref{CompFields_2_S} suggests that a further dressing operation aiming at erasing Lorentz symmetry is possible. In \cite{Attard-Francois2016_I} we showed that in the case of tractors, the vielbein $e={e^a}_\mu$ could be used to this purpose since it has the transformation $e^S=S\- e$, characteristic of a $\SO$-dressing field. This is the same process as in the  example of GR, treated in Section~\ref{From tetrad to metric formulation of General Relativity}.  The difference is that in GR one erases Lorentz symmetry and ends-up with ``nothing'', that is no gauge symmetry but only coordinates transformations characteristic of geometric objects living on $\M$, while in the tractor case one ends-up with Weyl rescalings as residual gauge symmetry in addition to coordinates transformations. Computing the residual Weyl symmetry after this second dressing displays a slightly different $C$-map to be used to perform the transformation of the composite fields, see \cite{Attard-Francois2016_I}. As a matter of fact,  in the literature two kinds of transformation law for tractors can be found, which in our framework  corresponds to either  erasing only the $K_1$-symmetry \cite{Gover-Shaukat-Waldron09, Gover-Shaukat-Waldron09-2}, or to erasing both $K_1$ and Lorentz-symmetries \cite{Bailey-et-al94, Curry-Gover2015}. 

Since there is no finite dimensional spin representation of $GL$, in the twistor case the vielbein cannot be used as  a second dressing field. By the way, looking at the $SL(2, \CC)$ gauge transformation of the vielbein, one sees that it is unsuited as a $\SL$-dressing field. So, as far as twistors are concerned, the process of symmetry reduction ends here.

\paragraph{BRST treatment}
The gauge group of the initial Cartan geometries are $\H$ and $\b\H$. The associated ghost $v\in\text{Lie}\H$ and  $\b v\in\text{Lie}\b\H$ split along the grading of $\LieH$ and $\b\LieH$,
\begin{align*}
v &=v_0+v_\iota
=v_\epsilon+v_\ss+v_\iota
\setlength\arraycolsep{4pt}
=\begin{pmatrix} \epsilon & \iota & 0 \\ 0 & s & \iota^t \\ 0 & 0 & -\epsilon \end{pmatrix},
\\
 \b v& =\b v_0+\b v_\iota 
 =\b v_\epsilon+\b v_\ss+\b v_\iota
 \setlength\arraycolsep{4pt}
=\begin{pmatrix} -(\b s^* - \sfrac{\epsilon}{2})  & -i \b \iota \\ 0 & \b s- \sfrac{\epsilon}{2} \end{pmatrix}. 
\end{align*} 
The BRST operator splits accordingly as $s=s_0+s_1= s_\ww+s_\l+s_1$. The algebra satisfied by  the gauge fields $\chi=\{ \varpi, \Omega, \vphi, \b\varpi, \b\Omega, \psi\}$, noted $\mathsf{BRST}$, is
\begin{align*}
s\varpi &=-Dv=-dv -[\varpi, v], 
&
s\Omega &=[\Omega, v], 
&
sv&=-v^2,
\\
s\b\varpi &=-D\b v=-d\b v -[\b\varpi, \b v], 
&
s\b\Omega &=[\b\Omega, \b v], 
&
s\b v&=-\b v^2,
\\
s\vphi &=-v\vphi ,
&
s\psi &=-\b v\psi ,
\end{align*}

From Section~\ref{Application to the BRST framework}, the composite fields $\chi_1=\{ \varpi_1,\Omega_1, \vphi_1,  \psi_1 \}$ satisfy a modified BRST algebra, formally similar but with composite ghost $v_1:=u_1\- v u_1 + u_1\-su_1$. From the finite gauge transformations of $u_1$, and the linearizations $\gamma_1\simeq\1 + v_\iota$ and $\sS \simeq\1 + v_\ss$, the BRST actions of $\K_1$ and $\SO$ are found to be:  $s_1u_1=-v_\iota u_1$ and $s_\l u_1=[u_1, v_\ss]$. This shows that the Lorentz sector is an instance of the general result \eqref{NewGhost2}. Using the linearizations $\sZ\simeq\1 + v_\epsilon$ and $k_1(z)\simeq\1 + \kappa_1(\epsilon)$, so that $C(z)=k_1(z)\sZ\simeq\1+c(\epsilon)=\1 + \kappa_1(\epsilon)+v_\epsilon$, the BRST action of $\W$ is $s_\ww u_1=-v_\epsilon u_1 + u_1 c(\epsilon)$. This shows that the Weyl sector is an instance of the general result \eqref{NewGhost3}. After a straightforward computation and a similar analysis for the complex case, we get the composite ghosts
\begin{align*}
v_1
&=c(\epsilon) + v_\ss
\setlength\arraycolsep{5pt}
=\begin{pmatrix} \epsilon & \d\epsilon & 0 \\ 0 & s & \d\epsilon^t \\ 0 & 0 & -\epsilon \end{pmatrix},
&
\b v_1 &=\b c(\epsilon) + \b v_\ss
\setlength\arraycolsep{5pt}
=\begin{pmatrix} -\left(\b s^* -\tfrac{\epsilon}{2}\1_2 \right) & -i\b{\d\epsilon} \\ 0 & \b  s - \tfrac{\epsilon}{2}\1_2 \end{pmatrix}, 
\end{align*}
where $\d\epsilon:=\d_a \epsilon=\d_\mu \epsilon\  {e^\mu}_a$. The ghost of conformal boosts, $\iota$, has disappeared from these new ghosts, replaced by the first derivative of the Weyl ghost. This means that $s_1 \chi_1=0$, which reflects the $\K_1$-gauge invariance of the composite fields $\chi_1$. The composite ghost $v_1$ only depends on $v_\ss$ and $\epsilon$: it encodes the residual $\K_0$-gauge symmetry. The  algebra satisfied by the composite fields $\chi_1$, denoted by $\mathsf{BRST}_{\ww, \l}$,  is then simply
\begin{align*}
s\varpi_1&=-D_1v_1=-dv_1 -[\varpi_1, v_1], 
&
s\Omega_1&=[\Omega_1, v_1], 
&
sv_1&=-v_1^2,
\\
s\b\varpi_1&=-D_1 \b v_1=-d \b v_1 -[\b\varpi_1, \b v_1], 
&
s\b\Omega_1&=[\b\Omega_1, \b v_1], 
&
s\b v_1&=-\b v_1^2,
\\
s\vphi_1&=-v_1\vphi_1,
&
s\psi_1&=-\b v\psi_1,
\end{align*}
and reproduces the infinitesimal version of \eqref{CompFields_1_S}, \eqref{CompFields_1_S_bar}, \eqref{CompFields_2_S}, \eqref{CompFields_2_S_bar} (Lorentz/$SL(2, \CC)$ sector) and \eqref{varpi_1_Z}, \eqref{varpi_1_Z_bar},  \eqref{Omega_1_Z},  \eqref{Tractor_Connection_1}  \eqref{Twistor_Connection_1} (Weyl sector).  Explicit results are obtained via simple matrix calculations, we refer to \cite{Attard-Francois2016_I, Attard-Francois2016_II} for all details. 

Since $v_1=c(\epsilon)+v_\ss$, $\mathsf{BRST}_{\ww, \l}$ splits naturally as Lorentz and Weyl subalgebras, $s=s_\ww+s_\l$. The Lorentz sector $(s_\l, v_\ss)$ shows the composite fields $\chi_1$ to be genuine Lorentz gauge fields. While the Weyl sector $(s_\ww, c(\epsilon))$ shows $\chi_1$ to be generalized Weyl gauge fields.

\subsection{Discussion}
\label{Discussion3}

Today, tractors and twistors are terms whose meaning extends beyond their original context of definition, conformal (and projective) geometry, and are quite broad concepts in the theory of parabolic geometries \cite{Cap-Slovak09}. In their original meaning, most often tractor and local twistor bundles are constructed in a ``bottom-up'' way, starting with a conformal manifold $(\M, c)$ and building a gauge structure on top of it.

 First, one poses a defining differential equation on $(\M, c)$. In the case of tractors, this is the \emph{almost Einstein equation} (AE)
\begin{align*}
\nabla_\mu \nabla_\nu \s - \sP_{\mu\nu} \s  - \tfrac{g_{\mu\nu}}{n} \left( \Delta\s - \sP \s \right)=0,
\end{align*}
with $\s$ a $1$-conformal density ($\h\s=z\- \s$), $\nabla$ the Levi-Civita connection associated to a choice of metric $g_{\mu\nu} \in c$,  $\Delta:=g^{\mu\nu}\nabla_\mu\nabla_\nu$, and $\sP:=g^{\mu\nu}\sP_{\mu\nu}$. For twistors, one  defines the \emph{twistor equation}
\begin{align*}
{\nabla^{(A}}_{A'}\omega^{B)}=0, \quad \text{ or equivalently }\quad  \nabla_{AA'}\omega^B  -  \tfrac{1}{2}\delta^B_A\ \nabla_{CA'}\omega^C=0,
\end{align*}
 where $\omega^B: \M \rarrow \CC^2$ is a Weyl spinor. Then one prolongs these equations, recast them as first order systems. These are interpreted as first order differential operators acting on multiplets: $\nabla^\T V=0$ and $\nabla^\sT Z=0$ respectively, where $V=(\s, \ell_\mu, \rho)$  and $Z=(\omega^A, \pi_{A'})$. The transformations of the components of $V$ and $Z$ under Weyl rescaling of the metric is given either by definition, when the components are functions of the metric ($V$), or by choice ($Z$). This takes some algebra to prove. With still more algebra, one shows that these transformation laws also apply to $\nabla^\T V$ and $\nabla^\sT Z$. But then $V$ and $Z$ are interpreted as parallel sections of some  vectors bundles over $\M$, the \emph{standard tractor bundle} $\T$ and \emph{local twistor bundle} $\sT$ respectively, which are endowed with their linear connections, the \emph{tractor connection} $\nabla^\T$ and \emph{twistor connection} $\nabla^\sT$ (hence the notation). Their commutators  $[\nabla^\T, \nabla^\T ]V=  \kappa V$ and  $[\nabla^\sT, \nabla^\sT]Z=  \mathsf{K} Z$ are said to define respectively the tractor and twistor curvatures.

Thus, starting from differential equations on $(\M, c)$, one ends-up with a gauge structure on top of it in the form of the tractor and twistor bundles and their connections. The latter provide natural conformally covariant calculi for torsion-free conformal manifolds. We refer the reader to \cite{Bailey-et-al94, Curry-Gover2015} for detailed calculations of this bottom-up procedure in the tractor case, and to the classic \cite[Sec.~6.9]{Penrose-Rindler-vol2} for the twistor case. See also  \cite[Sec.~6.1]{Bailey-Eastwood91}, which extends the twistor construction to paraconformal manifolds. It has been noticed that the tractor and twistor bundles can be seen as associated bundles of the principal Cartan bundle $\P(\M, H)$, and a link between the normal conformal Cartan connection and the twistor  $1$-form was drawn by Friedrich \cite{Friedrich77}. Nevertheless, the construction via prolongation has been deemed more explicit in \cite{Bailey-Eastwood91}, and more intuitive and direct in \cite{Bailey-et-al94}, than the viewpoint in terms of associated vector bundles.

\smallskip
However, our procedure present several advantages. Starting from a ``bigger'' gauge structure over $\M$ controlled by the conformal Cartan bundle $\P$ and a double cover complex version $\b\P$, we obtain the vectors bundles endowed with covariant derivatives  $(E_1, D_1)$ and $(\sE_1, \b D_1)$ in a very straightforward and systematic way by symmetry reduction. So, our constructive procedure via the dressing method is ``top-down'' and involves much less calculations than the usual ``bottom-up'' approach outlined above, and is arguably more direct and intuitive. 
 
Furthermore, these bundles reduce to the usual tractor and twistor bundles and their respective covariant derivatives when restricted to the \emph{normal} Cartan geometry, and one gets $(E_1, D_{\n, 1})=(\T, \nabla^\T)$ and  $(\sE_1, \b D_{\n, 1})=(\sT, \nabla^\sT)$. So, here we effortlessly generalize the tractor and twistor derivatives, providing essentially tractor and twistor calculi for conformal manifolds with torsion. It follows  that if $\varpi_{\n,1}$  and $\b\varpi_{\n, 1}$ are the genuine tractor and twistor $1$-forms, then $\varpi_1$ and $\b\varpi_1$ may be labeled as \emph{generalized tractor and twistor $1$-forms}. 

Our approach allows to clearly highlight the fact that, while tractors, twistors, and the associated (generalized) $1$-forms and curvatures are genuine Lorentz/$SL(2,\CC)$ gauge fields, they are gauge fields of generalized kind w.r.t. Weyl rescaling gauge symmetry, transforming using a $1$-$\alpha$-cocycle on the Weyl group. A fact that, as far as we know, has never been noticed. 

Let's finally notice that in this framework, one can easily write a Yang-Mills-type Weyl-invariant Lagrangian and compute the corresponding field equations. It turns out that this Lagrangian reproduces Weyl gravity if one restricts to a normal Cartan connection, as was shown in \cite{AFL2016_I}. This by the way explains the equivalence between the Bach equation and the Yang-Mills equation for the normal conformal Cartan connection \cite{Korz-Lewand-2003} or the twistor $1$-form \cite{Merkulov1984_I}.

\section{Conclusion} 
\label{Conclusion} 

The dressing field method of gauge symmetry reduction is a fourth way, beside gauge fixing, SSBM, and the bundle reduction theorem, to handle challenges one faces in gauge theories. As a matter of fact, as mentioned at the end of Section~\ref{Local construction and physics}, it is relevant in many places in gauge fields theories, from QCD to anomalies in QFT. In this review paper we have outlined the main general results of the method concerning the construction of partially gauge invariant composite fields out of the usual gauge variables, and discussed two important cases where their residual gauge transformations  can be treated on a general ground. Interestingly, we saw a case in which the composite fields are gauge fields of an unusual geometric nature, so that we label them ``generalized'' gauge fields.

We have shown that the method applies to the BEHGHK mechanism pivotal to the electroweak model. In doing so, we highlighted the fact that the notion of spontaneously broken gauge symmetry, which have long raised doubts among both philosophers of science and lattice gauge  theorists (in view of the Elitzur theorem), is dispensable and anyway unnecessary for the empirical success of the Standard Model. This result is thus   satisfying from a philosophical standpoint, and does not question the heuristic power of the gauge principle. 

We have argued that the usual switching between the tetrad and metric formulations of GR is a simple application of the dressing field method. In doing so, we have stressed that, contrary to what is sometimes said, the linear connection $\Gamma$ and the Lorentz connection $A$ are not mutual gauge transforms, even if one considers GR as a gauge theory on the frame bundle $L\M$. Actually, to recover $A$ from $\Gamma$ one needs the bundle reduction theorem, and to get  $\Gamma$ from $A$ one needs the dressing field method.  So that, in this instance, these are reciprocal operations.

The method applied to the conformal Cartan geometry and its spinorial version allows to obtain  generalizations of the tractor and twistor calculi for conformal manifolds, extending them to manifolds with torsion, in a very straightforward ``top-down'' way. It happens to be computationally much more economical than the usual ``bottom-up'' approach by prolongation of the Almost Einstein  and twistor equations, and arguably more direct and intuitive. Also, we have seen that tractors and  twistors, while being genuine Lorentz gauge fields, are generalized gauge fields as far as Weyl rescaling symmetry is concerned. 

\smallskip
One suspects that still more instances of the dressing field method could be found in the literature on gauge theories. Furthermore, its simplicity  may put within reach results otherwise difficult to obtain by other approaches; the example of tractor calculi for various parabolic geometries and their application to physics comes to mind. It is our hope that this approach could contribute to clarify and enrich some aspects of gauge field theories in physics. 

In its present form, the method relies on the defining (structural) relations for gauge transformations: as already mentioned, while the field contents are different, definitions \eqref{CompFields} look \emph{algebraically} like gauge transformations \eqref{ActiveGT}.\footnote{Let us mention here how it is has been difficult, in several occasions, to convince some colleagues that these relations are not mathematically on the same footing.} This is a key ingredient of the method. One can raise the question about some possible other routes one could elaborate to define dressed fields on which a part of the gauge symmetry is erased, but not using gauge transformation-like relations. 

Finally, to make the dressing field method a full-fledged approach to gauge QFT, the question of its compatibility with quantization must be addressed. In particular, do the operations of quantization and of reduction by dressing commute? So far, the question has not been fully addressed. One can find in \cite{Masson-Wallet} some hints that the problem is not easy and straightforward, mainly because we may first face the problem of the definition of a mathematically sound, let alone unique, quantization scheme. A rich topic in itself, that again exemplifies the fruitful cross-fertilization between physics and mathematics.

 \bibliography{Biblio-utf8}

\end{document}